# Image of the Kerr–Newman Black Hole Surrounded by a Thin Accretion Disk

Sen Guo[1], Yu-Xiang Huang[2,3], En-Wei Liang[2], Yu Liang[4], Qing-Quan Jiang[3], and Kai Lin[4]
[1] College of Physics and Electronic Engineering, Chongqing Normal University, Chongqing 401331, People's Republic of China
[2] Guangxi Key Laboratory for Relativistic Astrophysics, School of Physical Science and Technology, Guangxi University, Nanning 530004, People's Republic of China
[3] School of Physics and Astronomy, China West Normal University, Nanchong 637000, People's Republic of China
[4] Hubei Subsurface Multi-scale Imaging Key Laboratory, School of Geophysics and Geomatics, China University of Geosciences, Wuhan 430074, People's Republic of China

## Abstract

The image of a Kerr–Newman (KN) black hole (BH) surrounded by a thin accretion disk is derived. By employing elliptic integrals and ray-tracing methods, we analyze photon trajectories around the KN BH. At low observation inclination angles, the secondary image of particles is embedded within the primary image. However, as the inclination increases, the primary and secondary images separate, forming a hat-like structure. The spin and charge of the BH, along with the observer's inclination angle, affect the image's asymmetry and the distortion of the inner shadow. To investigate the redshift distribution on the accretion disk, we extended the inner boundary of the accretion disk to the event horizon. The results show that the redshift distribution is significantly influenced by the observation inclination angle. Furthermore, we conducted a detailed analysis of the KN BH image using fisheye camera ray-tracing techniques and found that the optical appearance and intensity distribution of the BH vary at different observation frequencies (specifically at 230 GHz and 86 GHz). We also examined differences in intensity distribution for prograde and retrograde accretion disk scenarios. Comparing observational at the two frequencies, we found that both the total intensity and peak intensity at 86 GHz are higher than those at 230 GHz.

*Unified Astronomy Thesaurus concepts:* Black hole physics (159); Kerr–Newman black holes (885); Black holes (162)

## 1. Introduction

Searching for direct evidence of black holes (BHs) in the Universe is one of the most intriguing topics in astronomical observations. By analyzing the trajectories of stars at the center of the Milky Way, astronomers have inferred the presence of a supermassive BH, known as Sagittarius A* (Sgr A*; R. Genzel et al. 1997). Cygnus X-1, one of the earliest confirmed X-ray binary star systems, has observational data indicating that its compact object has a mass approximately 14.8 times that of the Sun, suggesting that it is a BH (J. A. Orosz & C. D. Bailyn 1997). The merger of two compact celestial bodies, such as BHs or neutron stars, can release a powerful gravitational-wave signal from a BH merger. GW150914 was produced by the collision of two BHs with masses of 29 and 36 times that of the Sun, resulting in the formation of a larger BH with a mass of 62 times that of the Sun (B. P. Abbott et al. 2016). The Event Horizon Telescope (EHT) captures images of the supermassive BH at the center of the Messier 87* (M87*) galaxy, providing direct evidence of the surrounding structure of the BH. This image reveals a bright, circular structure (the glowing part of the accretion disk) and a dark region in the center, known as the BH's shadow (K. Akiyama et al. 2019a). In 2022, Akiyama et al. released an image of Sgr A* that not only revealed the existence of a supermassive BH at the center of the Milky Way galaxy but also confirmed the analysis of stellar trajectories (K. Akiyama et al. 2022a). These observational results not only validate the general relativity (GR) but also provide valuable data for investigating the physical properties of BHs and extreme environments present in the Universe.

The bright circular structure in the BH image is formed by the strong gravitational bending of light emitted by the material in the accretion disk surrounding the BH, while the dark region represents the BH's shadow. The fundamental orbits of photons and the spacetime structure around the BH determine the intrinsic characteristics of the dark region. In contrast, the characteristics of the bright region depend on the emission properties of the accretion disk in a real astronomical environment, including the accretion disk model and its underlying physical processes. Accretion disk models can theoretically predict the appearance of a BH. The earliest accretion disk model was proposed by Shakura and Sunyaev in 1973, describing the geometrically thin and optically thick standard accretion disk structure, known as the $\alpha$ disk (N. I. Shakura & R. A. Sunyaev 1973). The Shapiro–Lightman–Eardley model describes a geometrically thin and optically thin disk, accounting for relativistic effects that characterize the motion of matter near BHs (A. P. Lightman & D. M. Eardley 1974). The Novikov–Thorne model further extends this concept within the framework of GR, providing an example of a geometrically thin but optically thick relativistic accretion disk (V. P. Frolov & I. D. Novikov 1998). Using the Novikov–Thorne model, Luminet employed a semi-analytical method to derive the image of a Schwarzschild BH (J. P. Luminet 1979). Since then, the images of accretion disks in different gravitational spacetimes have been extensively investigated (T. Harko et al. 2009; Z. Kovacs et al. 2009; C. Bambi 2012; Z. Stuchlik & J. Schee 2014; B. Dănilă et al. 2015; J. Schee & Z. Stuchlik 2015; J. Dexter 2016; K. V. Staykov et al. 2016; G. Gyulchev et al. 2019, 2020; S. Paul et al. 2020; L. G. Collodel et al. 2021; G. Gyulchev et al. 2021;

S. Hu et al. 2022; I. I. Çimdiker et al. 2023; S. Guo et al. 2023a, 2023b; K. Hioki & U. Miyamoto 2023; Y.-X. Huang et al. 2023; M. Wang et al. 2023; P. Dyadina & N. Avdeev 2024; S. Hu et al. 2024).

Recently, an interesting model has proposed a detailed classification of BH shadows and rings within the framework of optically thin emission, demonstrating that the observational features of BHs are significantly influenced by the specific details of the accretion disk (S. E. Gralla et al. 2019). Another significant contribution of this work is the introduction of the concept of "critical curve," which coincides with the BH photon ring and that shadows appear within the critical curve. Inspired by this, Zeng et al. explored the influence of quintessence dark energy on BH photon rings, producing shadow images under different emission planes (X.-X. Zeng & H.-Q. Zhang 2020). The topic of "hairy" BH images is discussed in Q. Gan et al. (2021), where it is shown that the presence of hair leads to two unstable circular photon orbits within a certain range, forming two unstable photon spheres. Li et al. studied the possible optical appearance of $f(R)$ global monopole BHs and discovered that modified gravity can produce observational results similar to those of Schwarzschild BHs (G.-P. Li & K.-J. He 2021a). Many other excellent works in this field have also been published in recent years, contributing to a deeper understanding of BH imaging and related phenomena (X.-X. Zeng et al. 2020, 2022a, 2022b, 2023; G.-P. Li & K.-J. He 2021b; J. Peng et al. 2021; L. Chakhchi et al. 2022; Y. Guo & Y.-G. Miao 2022; S. Guo et al. 2022a, 2023c; S. Hu et al. 2022; H.-M. Wang et al. 2022; X.-J. Wang et al. 2023). Our previous work also examined the optical morphology of the Hayward BH and found that the contribution of direct radiation exceeded 95% (S. Guo et al. 2022b). The features displayed in the BH image include a central brightness depression and multiple concentric ring structures in the outer region. Note that these studies were conducted within a spherically symmetric BH background.

According to predictions from the EHT, both M87* and Sgr A* are likely rotating BHs (K. Akiyama et al. 2022b). Analyzing the null geodesics in the spacetime of rotating BH is essential for explaining the observed horizon scales of astrophysical BHs (S. E. Gralla & A. Lupsasca 2020a, 2020b). In addition to examining the intrinsic spacetime characteristics of the BH, it is also crucial to consider the influence of the accretion flow. However, obtaining detailed images of BHs remains challenging. Accretion disk models often simplify the actual disk, resulting in models that may lack completeness. Additionally, the highly turbulent and variable nature of the inner accretion flow in real astronomical environments leads to significant variability in observed emissions. Despite these limitations, simplified accretion models can still provide valuable predictions for the potential observational outcomes of astrophysical BHs. The simulated image of a Kerr BH surrounded by an optically thin disk reveals two features: a central brightness depression and a narrow, bright "photon ring" (A. Chael et al. 2021). Using the *Gyoto* code to simulate images of a thin equatorial disk around a Kerr BH, the shape of the photon ring can also be inferred from interferometric images (H. Paugnat et al. 2022). By developing an analytical model of a thick accretion disk with a stable axisymmetric accretion flow, researchers can simulate the image of a BH's photon ring (F. H. Vincent et al. 2022). In the context of a thin disk background, Wang et al. explored the images of Kerr–de Sitter BHs and examined the impact of the cosmological constant on image features (K. Wang et al. 2024). The observational appearance of a Kerr–Melvin BH illuminated by a thin accretion disk has also been investigated. This study also delves into how the internal shadow and critical curves of a Kerr–Melvin BH can be utilized to estimate the magnetic field surrounding the BH (Y. Hou et al. 2022b).

The no-hair theorem describes BH characteristics in terms of mass, spin, and charge. It is therefore natural to consider Kerr–Newman (KN) BHs in addition to the previously discussed Kerr BHs. Although the charge around astrophysical BHs is typically neutralized quickly by plasma, a residual charge may still persist in certain accretion flow scenarios (T. Damour et al. 1978). Meanwhile, observations of the supermassive BH at the center of the Milky Way suggest that Sgr A* may indeed possess a small charge (M. Zajaček et al. 2018; G. Bozzola & V. Paschalidis 2021; P. Kocherlakota et al. 2021). Studies of the KN BHs have shown that charge does indeed affect the shape of the shadow (N. Tsukamoto 2018). Y. Hou et al. (2022a) investigated the multilensing effect of KN BHs and examined how charge influences higher-order images. Relying on the elliptic integrals and Jacobi elliptic functions, Wang et al. derived the geodesics of the KN BH (C.-Y. Wang et al. 2022). However, these studies have not addressed the optical morphology of a KN BH surrounded by thin accretion disks. It is important to investigate whether the inner shadow and photon rings can be used to estimate the electric field outside the BH, as well as to understand the influence of charge on observable characteristics.

In this analysis, we derive the ray trajectory near a KN BH surrounded by a geometrically and optically thin accretion disk and investigate the potential observational characteristics of this BH using ray-tracing methods. The structure of this paper is as follows: Section 2 calculates the ray trajectory of the KN BH from null geodesics and provides insights into the lensing bands and particle orbits. Section 3 explores the redshift and intensity of the KN BH surrounded by a thin accretion disk, presenting images for both prograde and retrograde accretion disk scenarios. Finally, we conclude by summarizing the key findings of the paper.

## 2. Ray Tracing of a KN BH

In this section, we briefly review the dynamics of KN BHs and discuss ray tracing based on the trajectories of orbital particles around KN BHs. In the Boyer–Lindquist coordinate system, the KN BH is described as follows (J. M. Bardeen 1973):

$$ds^2 = -\left(1 - \frac{2Mr - Q^2}{\Sigma(r)}\right)dt^2 + \frac{\Sigma(r)}{\Delta(r)}dr^2 + \Sigma(r)d\theta^2 - \frac{2(2Mr - Q^2)a\sin^2\theta}{\Sigma(r)}dtd\phi + \left[\sin^2\theta(r^2 + a^2) + \frac{(2Mr - Q^2)a^2\sin^4\theta}{\Sigma(r)}\right]d\phi^2,$$
(1)

where

$$\Sigma(r) = r^2 + a^2\cos^2\theta, \quad \Delta(r) = r^2 + a^2 + Q^2 - 2Mr,$$
(2)

in which the parameters $M$, $Q$, and $a$ represent the BH mass, charge, and spin, respectively. It is important to note that the spin parameter is defined as $a = J/M$, where $J$ is the angular momentum of the rotating BH. The roots of $\Delta(r)$ correspond to the inner and outer horizons of the KN BH, i.e.,

$$r_H = M \pm \sqrt{M^2 - (Q^2 + a^2)}. \quad (3)$$

For the above equation to hold, the relationship between the charge $Q$ and the spin $a$ must satisfy

$$Q^2 + a^2 \leqslant M^2. \quad (4)$$

According to the observational results, astrophysical BHs can only carry a small amount of charge. Therefore, it is reasonable to assume that the charge range of KN BHs is $0 \leqslant Q \leqslant 0.5M$ (G. Bozzola & V. Paschalidis 2021; P. Kocherlakota et al. 2021). Furthermore, based on the description of Kerr BHs, the spin of the BH should be limited within the range of $0 \leqslant a \leqslant M$ (K. Akiyama et al. 2019a). When the charge approaches zero, the KN BH reduces to a Kerr BH. Similarly, when the angular momentum approaches zero, the metric reduces to that of a Reissner–Nordström BH.

### 2.1. Photon Trajectory Near a KN BH

To study the motion trajectory of photons around KN BHs, it is essential to analyze the evolution of particles in BH spacetime. The motion of a photon in the BH background is governed by the geodesic equation, which describes the path that a photon follows under the influence of the spacetime curvature, which is (J. M. Bardeen 1973)

$$\frac{d^2 x^\mu}{d\lambda^2} + \Gamma^\mu_{\alpha\beta} \frac{dx^\alpha}{d\lambda} \frac{dx^\beta}{d\lambda} = 0, \quad (5)$$

where $\lambda$ denotes an affine parameter and $\Gamma^\mu_{\alpha\beta}$ represents the Christoffel symbols of the background geometry. When combined with appropriate initial conditions, the geodesic Equation (5) can be solved for the given metric. However, this approach is often quite complex to handle. Instead, we can utilize the Hamilton–Jacobi formalism to solve the geodesics. In this spacetime, the Hamilton–Jacobi equation, expressed in terms of the metric tensor $g^{\mu\nu}$ is given by B. Carter (1968) and S. Chandrasekhar (1998) as

$$\frac{\partial S}{\partial \lambda} = -\frac{1}{2} g^{\mu\nu} \frac{\partial S}{\partial x^\mu} \frac{\partial S}{\partial x^\nu}. \quad (6)$$

Similar to the analysis of spherically symmetric BHs, photon trajectories can be described using conserved quantities. In the case of a rotating BH, the trajectory is characterized by three conserved quantities: $p_t = -E$, $p_\phi = L$, and $\mathcal{C} = p_\theta^2 - \cos^2\theta(a^2 p_t^2 - p_\phi^2 \csc^2\theta)$. These quantities correspond to the energy-rescaled angular momentum $\xi$ and the energy-rescaled Carter integral $\eta$, respectively, i.e., (B. Carter 1968)

$$\xi = \frac{L}{E}, \quad \eta = \frac{\mathcal{C}}{E^2}. \quad (7)$$

For KN BHs, the four-momentum $p^\mu$ of photons moving along their trajectories can be expressed as (C.-Y. Wang et al. 2022)

$$\frac{\Sigma}{E} p^r = \pm_r \sqrt{\mathcal{A}(r)}, \quad (8)$$

$$\frac{\Sigma}{E} p^\theta = \pm_\theta \sqrt{\mathcal{B}(\theta)}, \quad (9)$$

$$\frac{\Sigma}{E} p^\phi = \frac{\xi}{\sin^2\theta} + \frac{a}{\Delta(r)}(a^2 + r^2 - a\xi) - a, \quad (10)$$

$$\frac{\Sigma}{E} p^t = a(\xi - a\sin^2\theta) + \frac{a^2 + r^2}{\Delta(r)}(a^2 + r^2 - a\xi), \quad (11)$$

where the $\mathcal{A}(r)$ and $\mathcal{B}(\theta)$ are the radial and angular potentials, respectively, and their specific forms are

$$\mathcal{A}(r) = -\Delta(r)[(\xi - a)^2 + \eta] + (a^2 + r^2 - a\xi)^2, \quad (12)$$

$$\mathcal{B}(\theta) = a^2 \cos^2\theta + \eta - \xi^2 \cot^2\theta. \quad (13)$$

The symbol $\pm_r$ and $\pm_\theta$ represent the signs of $p^r$ and $p^\theta$, respectively, while $r$ and $\theta$ denote the changes in the ray trajectory at the radial and angular inflection points. The turning points in the radial and angular directions occur at the zeros of the radial potential $\mathcal{A}(r)$ and the angular potential $\mathcal{B}(\theta)$, respectively. Considering a scenario where light emitted from a source with coordinates $(t_s, r_s, \theta_s, \phi_s)$ reaches an observer at infinity with coordinates $(t_o, \infty, \theta_o, \phi_o)$, we introduce the Mino time $\tau$ to parameterize the trajectory. The trajectory equation in terms of Mino time is given by S. E. Gralla & A. Lupsasca (2020b) as

$$\frac{dx^\mu}{d\tau} \equiv \frac{\Sigma}{E} p^\mu. \quad (14)$$

Based on Equations (8)–(11), we can derive the integral form of the motion equation (S. E. Gralla & A. Lupsasca 2020b):

$$\Delta t = t_o - t_s = I_t + a^2 G_t, \quad (15)$$

$$\tau = I_r = G_\theta, \quad (16)$$

$$\Delta\phi = \phi_o - \phi_s = I_\phi + \xi G_\phi, \quad (17)$$

where

$$I_t = \fint_{r_s}^{r_o} \frac{(2Mr - Q^2)(a^2 + r^2 - a\xi) + \Delta(r)r^2}{\pm_r \Delta(r)\sqrt{\mathcal{A}(r)}} dr, \quad (18)$$

$$I_r = \fint_{r_s}^{r_o} \frac{dr}{\pm_r \sqrt{\mathcal{A}(r)}}, \quad (19)$$

$$I_\phi = \fint_{r_s}^{r_o} \frac{2aMr - a^2\xi - aQ^2}{\pm_r \Delta(r)\sqrt{\mathcal{A}(r)}} dr, \quad (20)$$

$$G_t = \fint_{\theta_s}^{\theta_o} \frac{\cos^2\theta}{\pm_\theta \sqrt{\mathcal{B}(r)}} d\theta, \quad (21)$$

$$G_\theta = \fint_{\theta_s}^{\theta_o} \frac{d\theta}{\pm_\theta \sqrt{\mathcal{B}(r)}}, \quad (22)$$

$$G_\phi = \fint_{\theta_s}^{\theta_o} \frac{\csc^2\theta}{\pm_\theta \sqrt{\mathcal{B}(r)}} d\theta. \quad (23)$$

Here, the symbol $\fint$ represents the integral along the path connecting $x_s^\mu$ and $x_o^\mu$. Thus, we consider geodesic integration from both radial and angular perspectives. For angular integration, $\eta$ must be positive, and the geodesic can be obtained at the inflection point $\theta_\pm$, which is located above and below the equatorial plane (S. E. Gralla & A. Lupsasca 2020a)

$$\theta_\pm = \arccos(\mp\sqrt{\omega_+}), \quad (24)$$

where

$$\omega_\pm = \frac{1}{2} - \frac{\eta + \xi^2}{2a^2} \pm \sqrt{\frac{\eta}{a^2} + \frac{1}{4}\left(1 - \frac{\eta + \xi^2}{a^2}\right)^2}. \quad (25)$$

According to the method proposed in S. E. Gralla & A. Lupsasca (2020a), we utilize elliptic integration so that the positive angle path integration $\eta$ can be represented by the number of turning points encountered along the trajectory $z$ (S. E. Gralla & A. Lupsasca 2020b)

$$G_\theta = \frac{1}{a\sqrt{-\omega_-}}[2zK \pm_s F_s \mp_o F_o], \quad (26)$$

$$G_\phi = \frac{1}{a\sqrt{-\omega_-}}[2z\Pi \pm_s \Pi_s \mp_o \Pi_o], \quad (27)$$

$$G_t = -\frac{2\omega_+}{a\sqrt{-\omega_-}}[2zE' \pm_s E'_s \mp_o E'_o], \quad (28)$$

where

$$K = K\left(\frac{w_+}{w_-}\right) = F\left(\frac{\pi}{2} \bigg\| \frac{w_+}{w_-}\right), \quad (29)$$

$$\Pi = \Pi\left(w_+ \bigg| \frac{w_+}{w_-}\right) = \Pi\left(w_+; \frac{\pi}{2} \bigg\| \frac{w_+}{w_-}\right), \quad (30)$$

$$E' = E'\left(\frac{w_+}{w_-}\right) = E'\left(\frac{\pi}{2} \bigg\| \frac{w_+}{w_-}\right), \quad (31)$$

$$F_i = F\left(\arcsin\left(\frac{\cos\theta_i}{\sqrt{w_+}}\right) \bigg\| \frac{w_+}{w_-}\right), \quad (32)$$

$$\Pi_i = \Pi\left(w_+; \arcsin\left(\frac{\cos\theta_i}{\sqrt{w_+}}\right) \bigg\| \frac{w_+}{w_-}\right), \quad (33)$$

$$E'_i = E'\left(\arcsin\left(\frac{\cos\theta_i}{\sqrt{w_+}}\right) \bigg\| \frac{w_+}{w_-}\right). \quad (34)$$

Here, the index $i$ can denote either the source ($i = s$) or observer ($i = o$). The parameters $K$, $\Pi$, and $E'$ represent the complete elliptic integrals of the first, second, and third kinds, respectively. Meanwhile, $F_i$, $\Pi_i$, and $E_i$ denote the incomplete elliptic integrals of the first, second, and third kinds, respectively.

For a KN BH, the radial potential is dependent on the BH charge $Q$. To find rays that correspond to two specific impact parameters, it is necessary to determine the roots of the radial potential based on Equations (2) and (12), which can be rewritten as

$$\mathcal{A}(r) = r^4 + Ur^2 + Or + D, \quad (35)$$

where

$$U = a^2 - \eta - \xi^2, \quad (36)$$

$$O = 2M\eta + 2M\xi^2 + 2a^2M - 4a\xi M, \quad (37)$$

$$D = -\eta a^2 - \eta Q^2 - \xi^2 Q^2 - a^2 Q^2 + 2a\xi Q^2. \quad (38)$$

According to Equation (35), there should be four roots that satisfy $r_1 + r_2 + r_3 + r_4 = 0$, taking the following form:

$$r_1 = -j - \sqrt{-\frac{U}{2} - j^2 + \frac{O}{4j}}, \quad (39)$$

$$r_2 = -j + \sqrt{-\frac{U}{2} - j^2 + \frac{O}{4j}}, \quad (40)$$

$$r_3 = j - \sqrt{-\frac{U}{2} - j^2 - \frac{O}{4j}}, \quad (41)$$

$$r_4 = j + \sqrt{-\frac{U}{2} - j^2 - \frac{O}{4j}}, \quad (42)$$

where

$$j = \sqrt{\frac{\varpi_+ + \varpi_- - \frac{U}{3}}{2}}, \quad (43)$$

$$\varpi_\pm = \sqrt[3]{-\frac{-\frac{U}{3}(\frac{U^2}{36} - D) - \frac{O^2}{8}}{2} \pm \sqrt{\frac{-\frac{U^2}{12} - D}{3}^3 + \frac{-\frac{U}{3}(\frac{U^2}{36} - D) - \frac{O^2}{8}}{2}^2}}. \quad (44)$$

Assuming the observer is at infinity, the photon should have a turning point at $r_4$. The unified expression for radial integration is then given by

$$I_i \sim \fint_{r_s}^{r_o} dr \sim \fint_{r_s}^{r_o} \cdots + 2\omega \fint_{r_t}^{r_s} dr \cdots. \quad (45)$$

In this way, we can derive the specific form of radial integration as

$$I_r = \int_{r_s}^{\infty} \frac{dr}{\sqrt{\mathcal{A}(r)}} + 2\omega \int_{r_4}^{r_s} \frac{dr}{\sqrt{\mathcal{A}(r)}}. \quad (46)$$

Due to the different light sources, radial integration should account for two distinct scenarios, i.e., (S. E. Gralla & A. Lupsasca 2020a)

$$I_r^{\text{total}} = \begin{cases} 2\int_{r_s}^{\infty} \frac{dr}{\sqrt{\mathcal{A}(r)}} & r_+ < r_4 \in \mathbb{R}, \\ \int_{r_+}^{\infty} \frac{dr}{\sqrt{\mathcal{A}(r)}} & \text{otherwise}, \end{cases} \quad (47)$$

where the root $r_4(\xi, \eta)$ must be a real number and greater than the horizon. Following the method for the Kerr BH (S. E. Gralla & A. Lupsasca 2020a), we also calculate the inverse form of radial integration by exchanging the positions of the source and the observer. In this scenario, $H_r$ becomes $-H_r$, and the source radius can be expressed as (S. E. Gralla & A. Lupsasca 2020a)

$$r_s = \frac{r_4 r_{31} - r_3 r_{41} sn^2\left(\frac{1}{2}\sqrt{r_{31}r_{42}} I_r - \mathcal{F}_o\big|_{r_{31}r_{42}}^{r_{32}r_{41}}\right)}{r_{31} - r_{41} sn^2\left(\frac{1}{2}\sqrt{r_{31}r_{42}} I_r - \mathcal{F}_o\big|_{r_{31}r_{42}}^{r_{32}r_{41}}\right)}, \quad (48)$$

where $\mathcal{F}_o$ is

$$\mathcal{F}_o = F\left(\arcsin\sqrt{\frac{r_{31}}{r_{41}}} \bigg| \frac{r_{32}r_{41}}{r_{31}r_{42}}\right). \quad (49)$$

Similar to spherically symmetric BHs, the geodesic trajectories (both radial and angular) of the KN BH are derived through the introduction of elliptic integrals. Similarly, photons orbiting around a KN BH can traverse the vicinity of the BH multiple times along their trajectory. Consequently, it is crucial to compute the fractional orbital number, termed as

$$n = \frac{G_\theta}{2\int_{\theta_-}^{\theta_+} d\theta \mathcal{B}(\theta)^{\frac{1}{2}}} = \frac{a\sqrt{-\omega_-}}{4K} I_r. \quad (50)$$

By using Equations (26) and (50), the relationship between orbital fraction $n$ and the number of turning points $z$ can be obtained as (S. E. Gralla & A. Lupsasca 2020b)

$$n = \frac{1}{2}z \pm_o \frac{1}{4}\left[(-1)^z \frac{F_s}{K} - \frac{F_o}{K}\right]. \quad (51)$$

For the general case involving two impact parameters $\xi$ and $\eta$, the radial potential (Equation (12)) has four different roots (Equations (39)–(42)), where the real subset corresponds to the radial inflection point. Analogous to Schwarzschild BHs, a critical value of the impact parameter signifies the position where the photon ring of the BHs emerges. In the case of a rotating BH, the radial potential can exhibit a double root at a specific radius below the critical values denoted as $\tilde{\xi}$ and $\tilde{\eta}$, and $\mathcal{A}(\tilde{r}) = \mathcal{A}'(\tilde{r}) = 0$. We have

$$\tilde{\xi} = a + \frac{\tilde{r}(2(a^2 + Q^2)) - 3M\tilde{r}^2 + \tilde{r}^3}{a(M - \tilde{r})}, \quad (52)$$

$$\tilde{\eta} = -\frac{\tilde{r}^2(4a^2(Q^2 - M\tilde{r}) + (2Q^2 + \tilde{r}(\tilde{r} - 3M))^2)}{a^2(M - \tilde{r})^2}. \quad (53)$$

When a geodesic passes through the equatorial plane, its (squared) instantaneous momentum perpendicular to the equatorial plane, i.e., $p_\theta^2$ equals $\mathcal{B}(\pi/2) = \eta$. There is a requirement that $\tilde{\eta} > 0$ for geodesics intersecting the equatorial plane, which further excludes the situation known as "vortical" geodesics ($\tilde{\eta} < 0$). In the case of a Kerr BH, where $Q = 0$, this equation simplifies to a cubic equation, and solutions can be obtained in two directions through trigonometric functions (S. E. Gralla & A. Lupsasca 2020a):

$$\tilde{r}_\pm = 2M\left[1 + \cos\left(\frac{2}{3}\arccos\left(\pm\frac{a}{M}\right)\right)\right]. \quad (54)$$

Next, we will focus on Equation (53). For a KN BH, the expression for $\tilde{\eta}$ is a fourth-order function with respect to $\tilde{r}$, making the method of trigonometric function transformation ineffective. Consequently, we employed a numerical solution approach to address the equation, ensuring the accuracy of the critical curve for the KN BH. Note that the $\eta > 0$ for geodesics intersecting the equatorial plane. Simplification of Equation (53) leads to

$$-\tilde{r}^4 + 6M\tilde{r}^3 - (9M + 4Q^2)\tilde{r}^2 + (4a^2M + 12MQ^2)\tilde{r} - 4(a^2 + Q^2)Q^2 = 0. \quad (55)$$

Figure 1 illustrates the relationship between $\tilde{r}$ and $\tilde{\eta}$. Across various BH charges, the four roots of Equation (55) remain, preserving the order $\tilde{r}_1 < \tilde{r}_2 < \tilde{r}_3 < \tilde{r}_4$. Considering that the event horizon of the KN BH satisfies $\tilde{r}_1 < \tilde{r}_2 < r_H < \tilde{r}_3 < \tilde{r}_4$, it is necessary to designate $\tilde{r}_-$ and $\tilde{r}_+$ as $\tilde{r}_3$ and $\tilde{r}_4$, respectively.

### 2.2. Lensing Bands and Orbits of a KN BH

Assuming the observer is located at infinity, the orthogonal parameters of the photon reaching the observer are denoted as $(\alpha, \beta)$. The observation plane can be considered as a Cartesian coordinate plane, where the coordinates are proportionate to the cosine of the observer's direction in the sky. In our previous research (S. Guo et al. 2023a), we extensively discussed the projection method for spherically symmetric BHs. However, unlike our previous discussions, the KN BH involves two conserved quantities of photons. Hence, the expressions for $\alpha$ and $\beta$ are reformulated as described in S. E. Gralla & A. Lupsasca (2020a, 2020b) as

$$\alpha = -\frac{\xi}{\sin\theta_0}, \quad (56)$$

$$\beta = \pm_o\sqrt{\mathcal{B}(\theta_0)} = \pm_o\sqrt{\eta + a^2\cos^2\theta_0 - \xi^2\cot^2\theta_0}, \quad (57)$$

where $\theta_0$ represents the inclination angle of observation. Based on the aforementioned equations, the lensing band formed by photons with critical impact parameters ($[\tilde{\xi}(\tilde{r}), \tilde{\eta}(\tilde{r})]$) around the KN BH can be derived and expressed in terms of a set of coordinates ($[\tilde{\alpha}(\tilde{r}), \tilde{\beta}(\tilde{r})]$) on the observer's screen.

Figure 2 displays the lensing bands of the KN BH at different observation inclination angles, with a screen resolution of $2048 \times 2048$ pixels. The critical curves are shown as solid white lines. Three distinct observation inclination angles are selected, with one angle, $\theta_0 = 17$ deg, corresponding to the EHT's imaging angle for M87*. Upon comparison, it becomes apparent that as the observation inclination angle increases, the critical curve gradually extends toward the southwest ($\theta_0 < \pi/2$). As the inclination angle continues to increase and approaches the reversal angle, the critical curve nearly reverses its direction.

Continuing the discussion on the orbits of particles around the KN BH, we focus on equatorial sources for simplicity. Nonequatorial sources will be addressed in our forthcoming work. In this scenario, the observer is positioned at a specific observation inclination angle ($\theta_0 \neq 0$). As a result, the first kind of incomplete elliptic integral, $F_s$, becomes zero, leading to the following expression (S. E. Gralla & A. Lupsasca 2020a):

$$\sqrt{-\omega_-a^2}I_r + \text{sign}(\beta)F_o = 2zK, \quad (58)$$

where $F_o$ represents the first kind of incomplete elliptic integral, and $K$ denotes the first kind of complete elliptic integral. The equation provides the corresponding coordinates $(\alpha, \beta)$ for various radial positions $r_s$ and source orders $z$. These coordinates can be used to generate contour maps of particle trajectories. It is important to note that while we focus on radial behavior in this discussion, the same approach can be applied to derive angular behavior as well.

In Figure 3, the particle orbits around the KN BH are illustrated. When the observation inclination angle is small, the secondary image is nested within the primary image, similar to the behavior seen around a spherically symmetric BH. However, as the observation inclination angle increases, the primary and secondary images gradually separate, taking on a hat-like shape. Unlike spherically symmetric BHs, the BH's spin introduces an asymmetry between the first and second-order images in the east–west direction. It is worth noting that the charge has a minimal impact on the orbits, whereas the BH's spin and observation inclination angle have significant effects, especially in the inner shadow, where considerable distortion can occur.

By analyzing the rays on the $(\alpha, \beta)$ plane, we can create a schematic diagram illustrating plane ray tracing for various incident angles. The entire plane is divided into 50 regions, with light in each region being predominantly influenced by the strong gravitational field of the BH. As shown in Figure 4, the effect of the BH's charges on light is minimal, while the spin effect is significant, consistent with the observations from Figure 3. It is important to note that this form of planar ray tracing does not represent the final image of a BH but serves as a simplified depiction of light trajectories within the context of spacetime curvature. This approach provides a clearer illustration of the path of light around the KN BH.

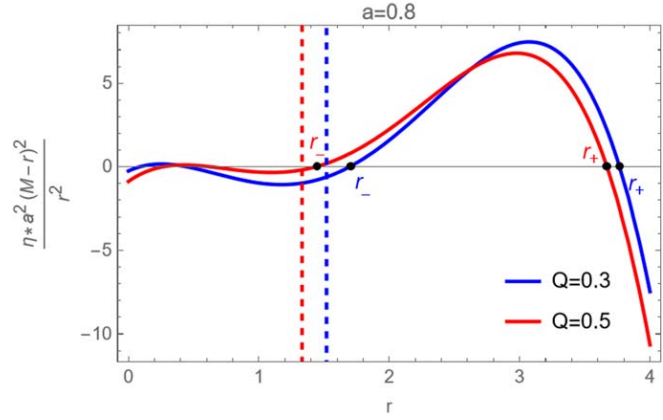

**Figure 1.** $\tilde{r}$ is depicted as a function of $\tilde{\eta}$. The intersection points of the solid lines with the *x*-axis represent the positions of the roots. The solid blue line corresponds to $Q = 0.3$, while the solid red line indicates $Q = 0.5$. The dashed line shows the location of the event horizon. The mass of the KN BH is set to $M = 1$, and the spin is $a = 0.8$.

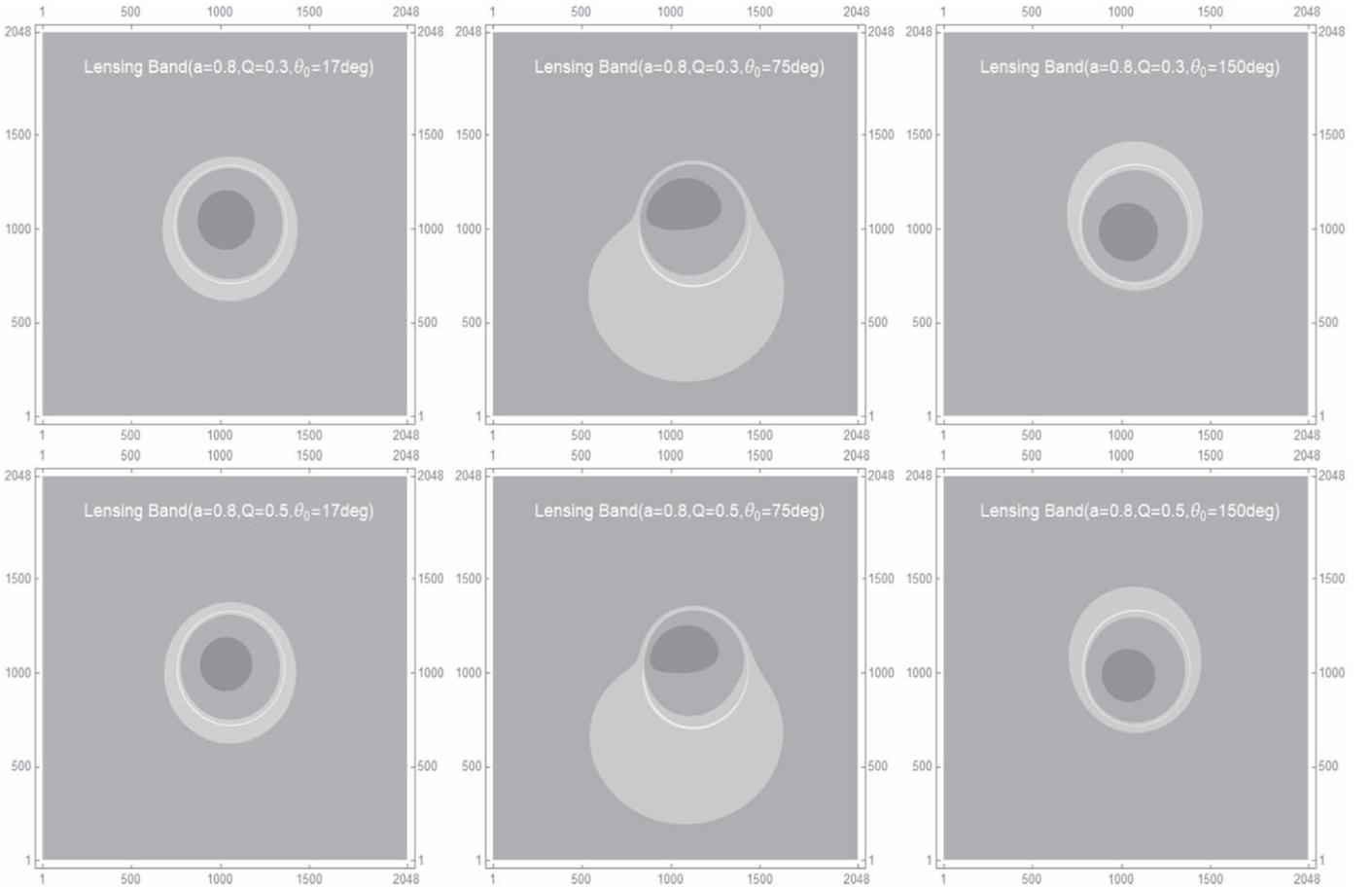

**Figure 2.** The lensing bands of the KN BH are shown under several representative BH parameters and observation angles. Top panel: BH spin is set to $a = 0.8$ and charge to $Q = 0.3$. Bottom panel: the BH spin remains $a = 0.8$, but the charge is increased to $Q = 0.5$. From left to right, the observation inclination angles are set to 17 deg, 75 deg, and 150 deg. The dark gray and light gray regions represent light rays that intersect the null geodesic and the equatorial plane of the BH once and twice, respectively. The critical curve is depicted as a white line, and the innermost region corresponds to the "inner shadow" of a KN BH. The mass of BH is set to $M = 1$.

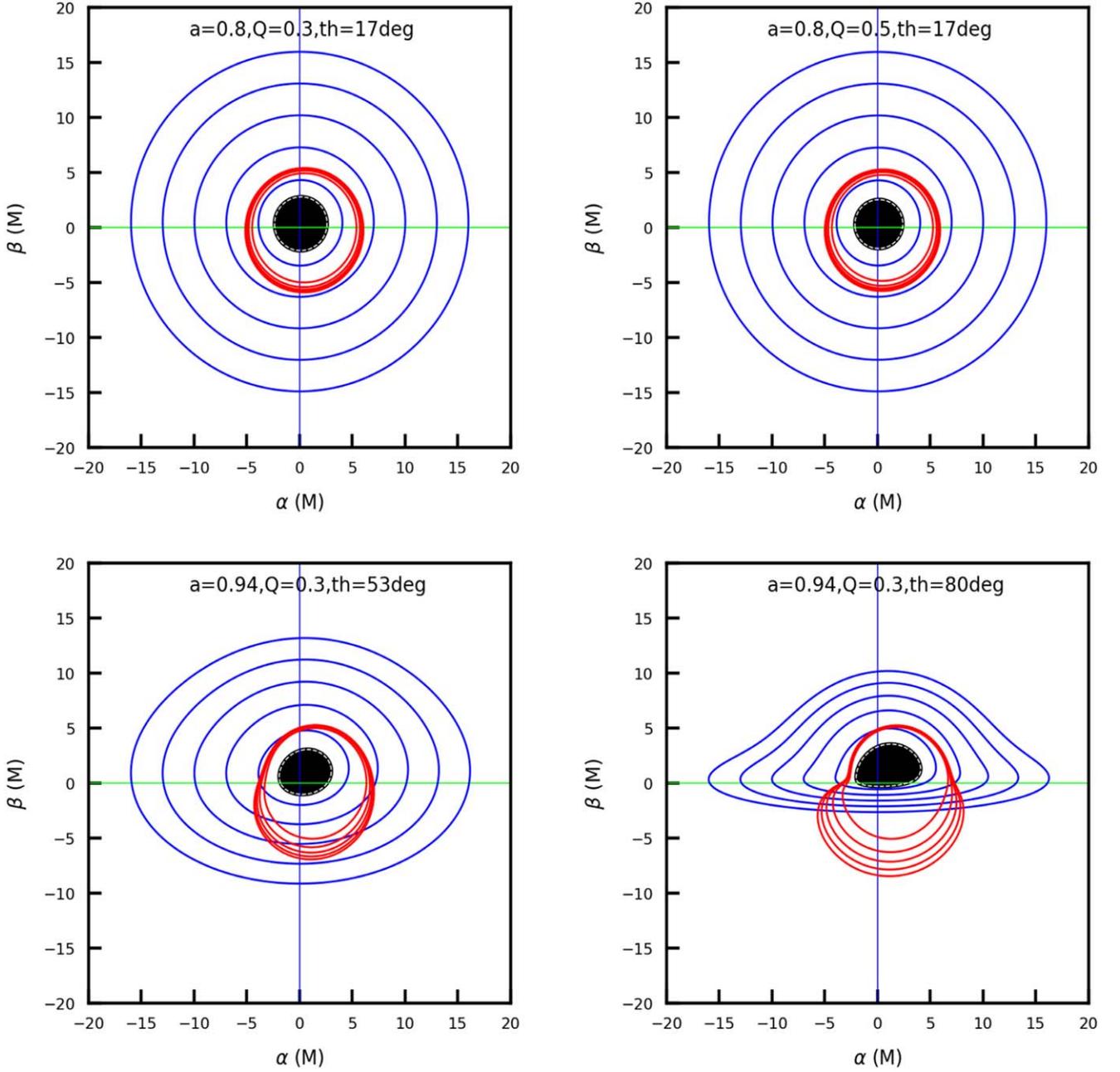

**Figure 3.** The direct (blue lines) and secondary (red lines) orbits of the KN BH are shown under several representative BH parameters and observation inclination angles. The mass of the BH is set to $M = 1$.

In this section, we analyzed null geodesics around a rotating charged BH and derived the critical curve of the KN BH using numerical methods. Additionally, we illustrated the motion trajectories of particles around the target BH and provided a simplified depiction of photon paths through planar ray tracing. These efforts serve as adequate preparation for the subsequent step of the KN BH image.

### 3. Accretion Disk and Image of a KN BH

Next, we will conduct a thorough investigation into the optical appearance of a KN BH surrounded by a thin accretion disk. Consider an accretion disk lying in the equatorial plane, characterized by geometric thin and optically thin properties. In traditional BH accretion disk models, the innermost stable circular orbit (ISCO) is typically regarded as the disk's inner boundary. The ISCO marks the outermost orbit where test particles in proximity to the BH become unstable due to inward radial perturbations.

In our analysis, we expanded upon the classical accretion disk model, where the velocity of test particles near the BH increases rapidly as their radial position approaches the ISCO. As a result, the inner portion of the accretion disk extends inward, effectively stretching closer to the BH's horizon. In simpler terms, the innermost section of the accretion disk now reaches nearer to the horizon. Radially, the outer boundary of the accretion disk extends beyond the ISCO, while the inner region pushes toward the BH's horizon. This updated model introduces two significant modifications: (1) Within the ISCO region, it becomes necessary to reconsider the effects of gravitational redshift and radiation transfer; (2) Ray-tracing techniques require a larger integration area. While this accretion disk concept was briefly mentioned in our prior work (S. Guo et al. 2022a), here we not only extend its application to KN BHs but also provide a more detailed derivation of the resulting images of BH accretion disks in both prograde and retrograde directions.

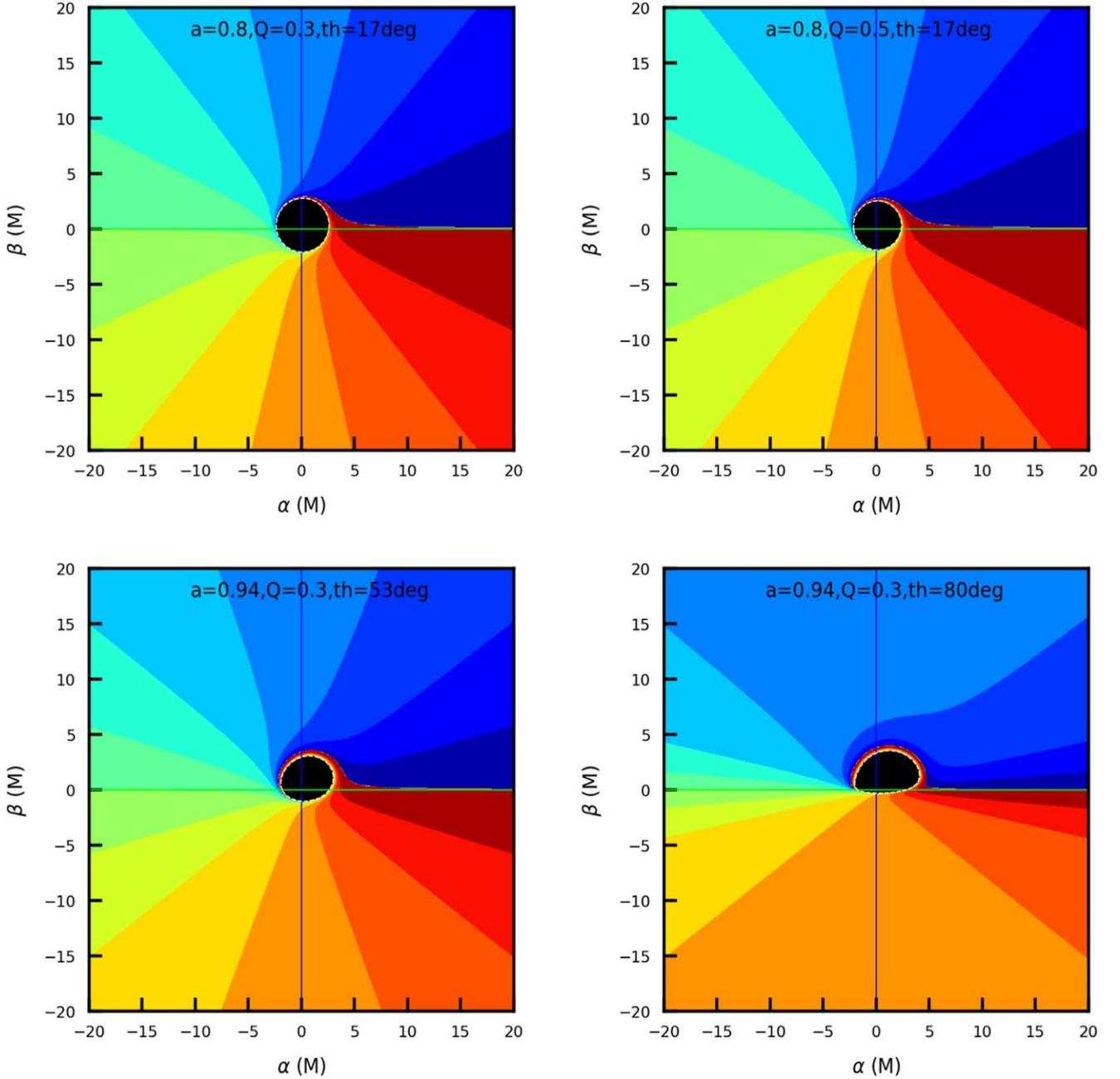

**Figure 4.** Ray trajectories around KN BHs on the ($\alpha$, $\beta$) plane. Set the mass of the BH to $M = 1$.

From the preceding discussion, it is clear that the ISCO marks the boundary for particle motion. Beyond this boundary, particles orbiting the accretion disk remain in a stable circular orbit, while within this boundary, they enter a plunging orbit. Following the normalization conditions of energy and angular momentum, the radial equation governing particle motion is (Y. Hou et al. 2022b)

$$u^r = -\sqrt{-\frac{V(r, E, L)}{g^{rr}}}, \qquad (59)$$

where

$$V(r, E, L) = (g^{tt}E^2 + g^{\phi\phi}L^2 - 2g^{t\phi}EL + 1)|_{\theta=\frac{\pi}{2}}. \qquad (60)$$

Based on the aforementioned equation, we derived the effective potential function of the KN BH. As illustrated in Figure 5, an increase in charge lowers the peak of the effective potential, while simultaneously causing the location of this peak to shift outward. Similarly, the spin of a BH produces a comparable effect, indicating that changes in these two parameters alter the BH's spacetime configuration and, consequently, influence the trajectories of particles in the vicinity of the KN BH.

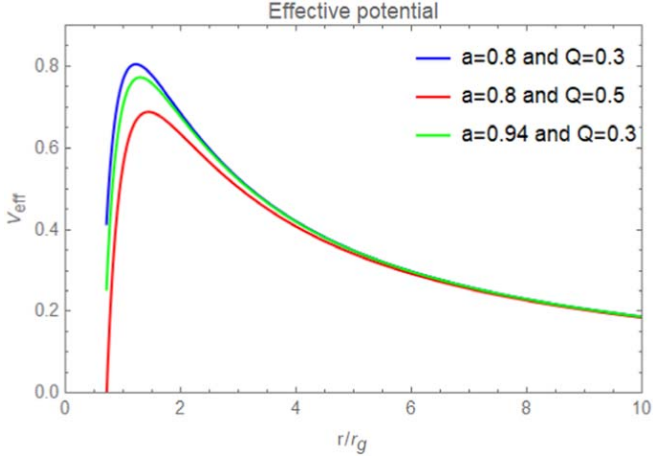

**Figure 5.** The effective potential function of the KN BH is determined by setting the BH's mass to $M = 1$.

Using the critical condition of the effective potential, $V = \partial_r V = 0$, we can derive the ISCO for particle motion near the KN BH. When $r > r_{\rm ISCO}$, the particle motion is governed by $V = \partial_r V = 0$. However, when $r < r_{\rm ISCO}$, the particle motion is described by the following equation (A. Chael et al. 2021):

$$u_c^r = -\sqrt{-\frac{V(r, E_{\rm ISCO}, L_{\rm ISCO})}{g^{rr}}}\bigg|_{\theta=\frac{\pi}{2}}. \quad (61)$$

Thus, we have obtained the separate motion behaviors of particles on either side of the ISCO.

### 3.1. Zero-angular-momentum Observer

Next, we will establish a fisheye camera model. To describe the local frame of the observer, we can select a standard orthogonal tetrad, denoted as (Z. Hu et al. 2021)

$$e_t = \delta \partial_t + \chi \partial_\phi, \quad e_r = \frac{1}{\sqrt{g_{rr}}} \partial_r, \quad (62)$$

$$e_\theta = \frac{1}{\sqrt{g_{\theta\theta}}} \partial_\theta, \quad e_\phi = \frac{1}{\sqrt{g_{\phi\phi}}} \partial_\phi, \quad (63)$$

where

$$\delta = \sqrt{\frac{g_{\phi\phi}}{g_{t\phi}^2 - g_{tt}g_{\phi\phi}}}, \quad \chi = -\frac{g_{t\phi}}{g_{\phi\phi}}\sqrt{\frac{g_{\phi\phi}}{g_{t\phi}^2 - g_{tt}g_{\phi\phi}}}. \quad (64)$$

Within the framework of a zero-angular-momentum observer, the photon's trajectory is reversible, making it reasonable to assume that any photon reaching the observer's position can be fully captured. The four-momentum of a photon, as measured by the zero-angular-momentum observer, is expressed as $p^\mu = \eta^{\mu\nu} e_\nu^\xi k_\xi$, with its component forms given by (K. Wang et al. 2024)

$$P^t = E(\delta - \chi\xi), \quad P^r = E\frac{1}{\sqrt{g_{rr}}}\frac{\pm_r \sqrt{\mathcal{A}(r)}}{\Delta_r}, \quad (65)$$

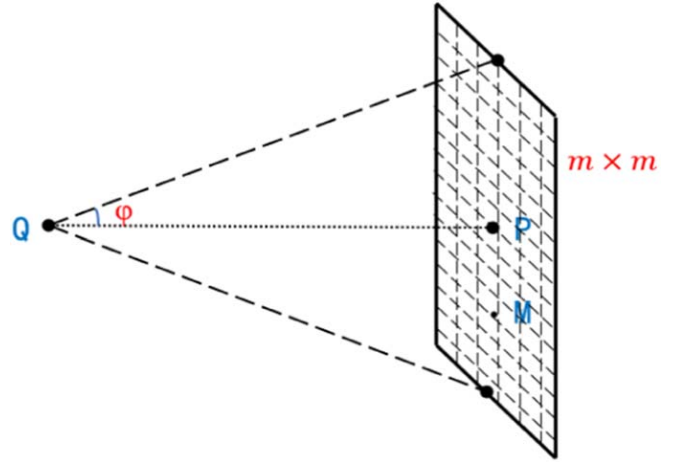

**Figure 6.** Fisheye camera field of view with $m \times m$ pixels.

$$P^\theta = E\frac{1}{\sqrt{g_{\theta\theta}}}\frac{\pm_\theta \sqrt{\mathcal{B}(\theta)}}{\Delta_\theta}, \quad P^\phi = E\frac{\xi}{\sqrt{g_{\phi\phi}}}. \quad (66)$$

Following the approach outlined in Z. Hu et al. (2021), the celestial coordinates $\Theta$ and $\Psi$ can be defined to label each photon ray. The relationship between these coordinates and the photon's four-momentum can be expressed as

$$x = -2\tan\frac{\Theta}{2}\sin\Psi, \quad y = -2\tan\frac{\Theta}{2}\cos\Psi. \quad (67)$$

By setting the initial position for integration at $(0, r_o, \theta_o, 0)$, the complete ray trajectory using the Hamiltonian regularization equation for null geodesics. After determining the complete ray trajectory, the next step is to consider how to capture images of the KN BH. One straightforward approach is to employ a pinhole camera for perspective projection. This model is simple and aligns with practical imaging principles, but it has the limitation of a relatively narrow field of view. This imaging technique is commonly referred to as the fisheye camera model (Z. Hu et al. 2021). We establish standard Cartesian coordinates on the imaging plane, with the center point labeled as $P$ and the observer located at another point $Q$ in space. The angle between $Q$ and the projection screen is defined as the field-of-view angle ($\varphi$). As shown in Figure 6, for any pixel point $M$ on the screen, the length of the screen can be expressed as (Z. Hu et al. 2021)

$$L = 2|\overrightarrow{QM}|\tan\frac{\varphi}{2}. \quad (68)$$

More precisely, the screen can be divided into $m \times m$ grids, similar to the pixel grids employed in typical camera imaging. Each grid has a length of

$$\mathcal{L} = 2\frac{|\overrightarrow{QM}|}{m}\tan\frac{\varphi}{2}. \quad (69)$$

Thus, for any pixel located at coordinates $(p, q)$, the total number of visible grids is given by

$$x = \mathcal{L}\left(p - \frac{m+1}{2}\right), \quad y = \mathcal{L}\left(q - \frac{m+1}{2}\right). \quad (70)$$

The effectiveness of this method is demonstrated by analyzing KN BH shadows using a fisheye camera. A screen with dimensions $2048 \times 2048$ pixels is employed to observe the image at infinity, as shown in Figure 7. To enhance the clarity of light trajectories in different regions, the screen is divided into four distinct quadrants. It is observed that an increase in charge results in significant bending of light around the KN BH, although this effect is relatively minor compared to the optical distortion caused by variations in the BH's spin. For a clearer view of the changes in the inner shadow, the central portion of Figure 7 is extracted and shown separately in Figure 8, with results consistent with our previous discussion.

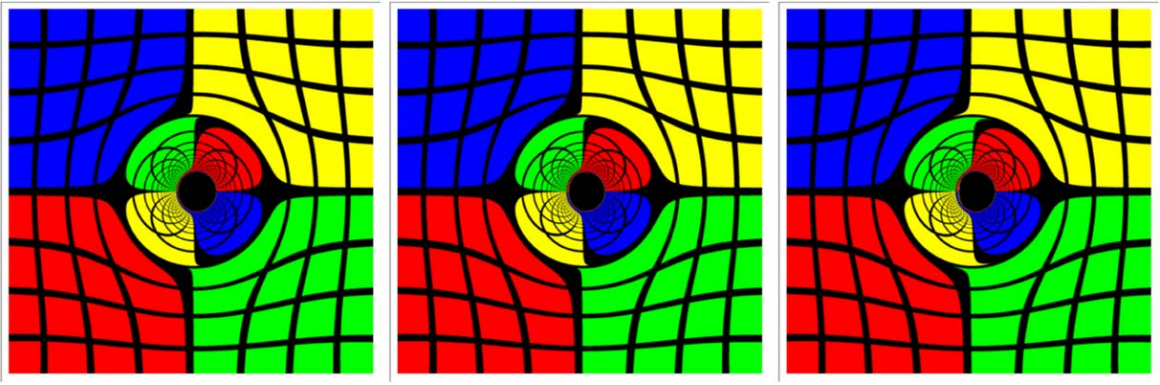

**Figure 7.** The shadow of the KN BH under several representative BH parameters using the numerical ray-tracing method. Left panel: BH spin $a = 0.8$ and charge $Q = 0.3$. Middle panel: BH spin $a = 0.8$ and charge $Q = 0.5$. Right panel: BH spin $a = 0.94$ and charge $Q = 0.3$. The mass of the BH is set to $M = 1$.

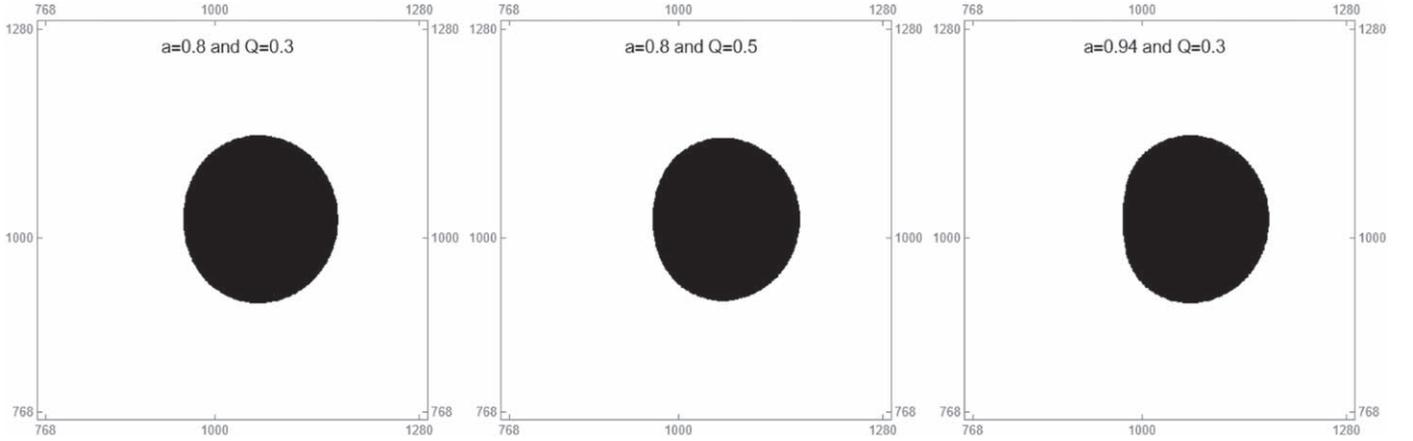

**Figure 8.** Inner shadow of the KN BH with the BH mass set to $M = 1$.

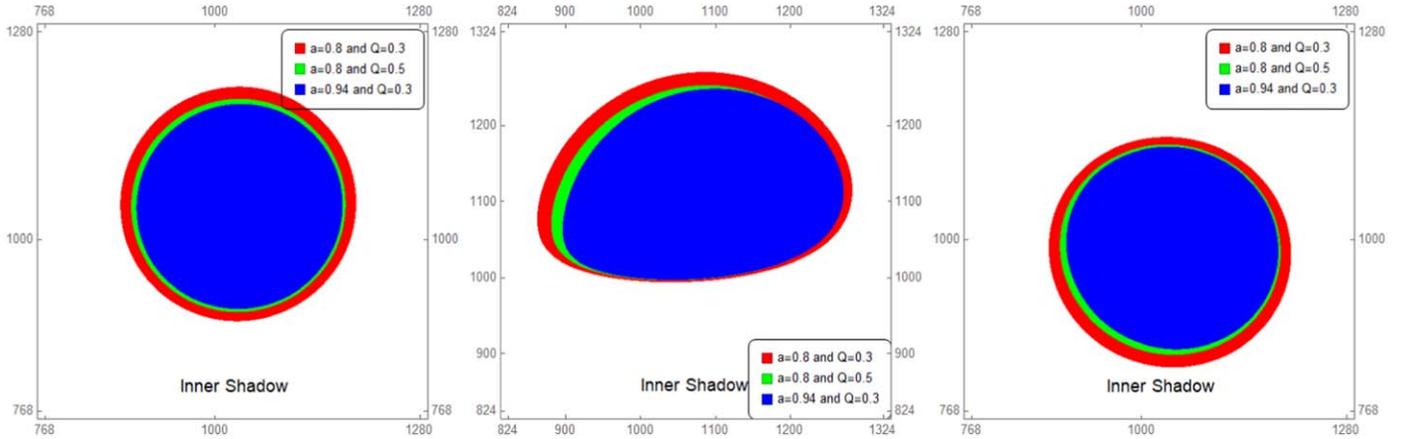

**Figure 9.** Inner shadow of the KN BH under several representative observation inclination angles. Left panel: observation angle $\theta_0 = 17$ deg. Middle panel: observation angle $\theta_0 = 75$ deg. Right panel: Observation angle $\theta_0 = 150$ deg. The mass of the BH is set to $M = 1$.

Additionally, varying the observation inclination angles allows for further comparison of the effects of different charges, observation inclination angles, and BH spins on the inner shadow of the KN BH, as illustrated in Figure 9.

### 3.2. Intensity and Redshift

In studying the light emitted from the accretion disk around a BH as it reaches the observer's plane, it is crucial to account for changes in light intensity due to divergence, absorption,

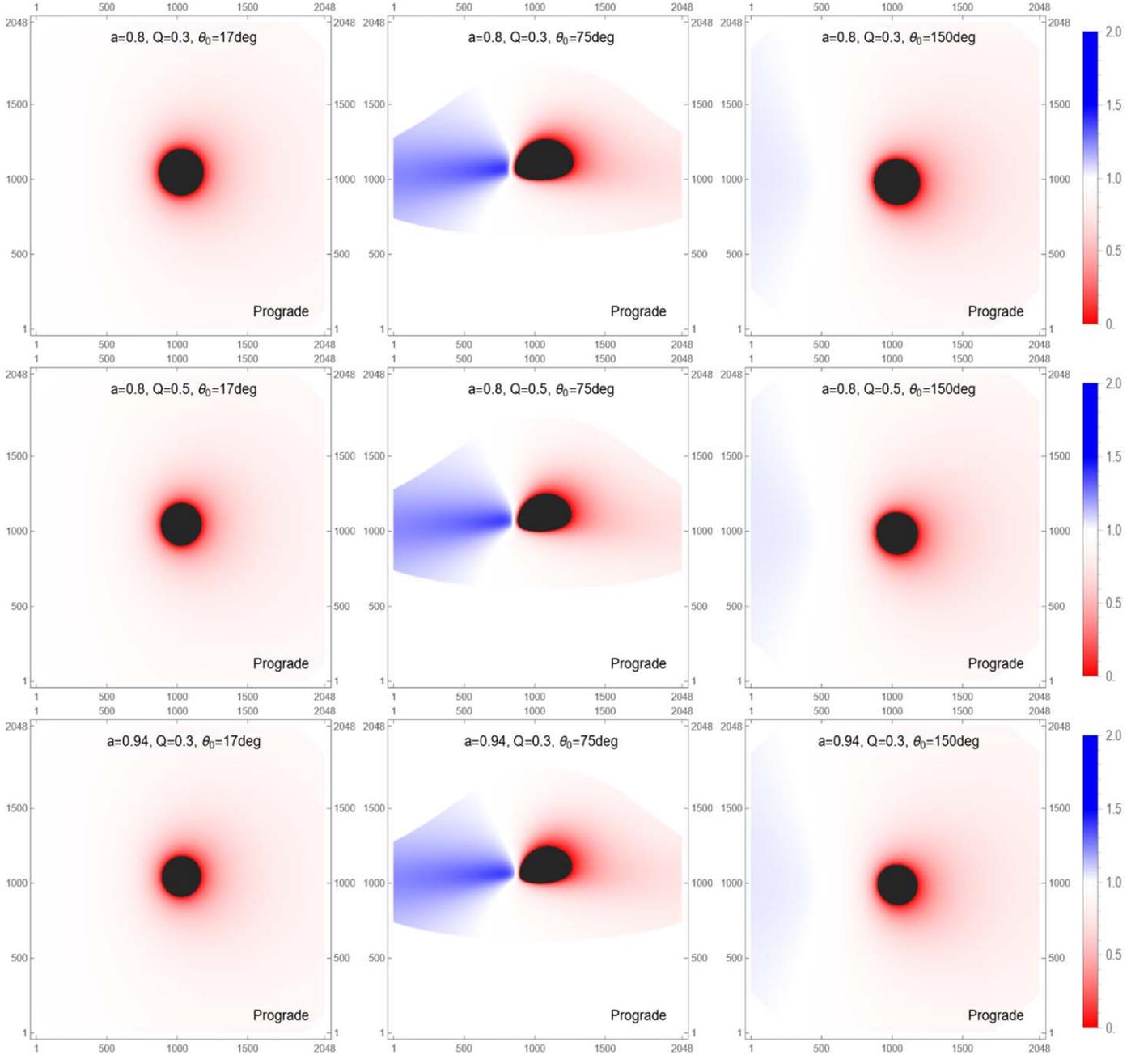

**Figure 10.** The redshift distribution of direct images under prograde disk. Left panel: observation angle $\theta_0 = 17$ deg. Middle panel: observation angle $\theta_0 = 75$ deg. Right panel: observation angle $\theta_0 = 150$ deg. From top to bottom, the spin of the KN BH is $a = 0.8$, $a = 0.8$, and $a = 0.94$, with a charge of $Q = 0.3$, $Q = 0.5$, and $Q = 0.3$. Set the mass of the BH to $M = 1$.

Doppler effects, and gravitational redshift. For simplicity, we assume that the accretion disk medium has negligible refractive effects. The intensity can be expressed as (R. W. Lindquist 1966)

$$\frac{d}{d\lambda}\left(\frac{I_\nu}{\nu^3}\right) = \frac{J_\nu - \kappa_\nu I_\nu}{\nu^2}, \quad (71)$$

where $\lambda$ serves as an affine parameter, $I_\nu$, $J_\nu$, and $\kappa_\nu$ denote the specific intensity, emissivity, and absorption coefficient at the frequency $\nu$, respectively. Since light propagates in a vacuum, both $J_\nu$ and $\kappa_\nu$ are zero. Consequently, the quantity $I_\nu/\nu^3$ remains conserved along the geodesic path.

From the previous discussion, it is evident that the accretion disk model we are considering is geometrically and optically thin and uniform. In the region outside the disk, both the emission and absorption coefficients are zero. Within the accretion disk surrounding the KN BH, these coefficients are treated as constants (a more detailed derivation can be found in Appendix B of Y. Hou et al. 2022b). By employing Equation (71), we can integrate the ray trajectory to determine the intensity at each position on the observer's screen. Consequently, we obtain (Y. Hou et al. 2022b)

$$I_{\nu_o} = \sum_{m=1}^{N_{max}} \left(\frac{\nu_o}{\nu_m}\right)^3 \frac{J_m}{\tau_{m-1}} \left[\frac{1 - e^{-\kappa_m f_m}}{\kappa_m}\right], \quad (72)$$

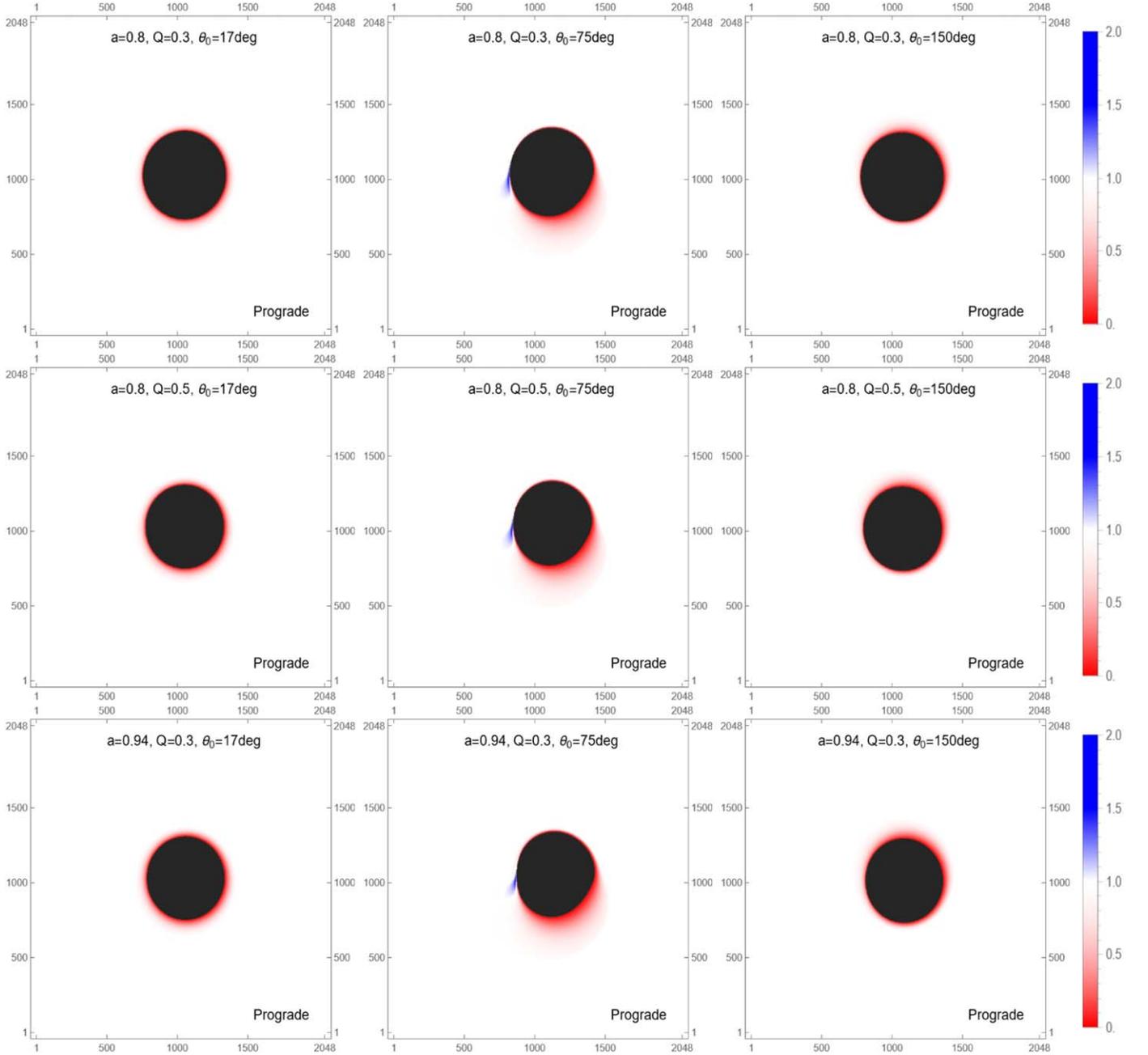

**Figure 11.** The redshift distribution of lensing images under prograde disk. Left panel: observation angle $\theta_0 = 17$ deg. Middle panel: observation angle $\theta_0 = 75$ deg. Right panel: observation angle $\theta_0 = 150$ deg. From top to bottom, the spin of the KN BH is $a = 0.8$, $a = 0.8$, and $a = 0.94$, with a charge of $Q = 0.3$, $Q = 0.5$, and $Q = 0.3$. Set the mass of the BH to $M = 1$.

where $\nu_o$ denotes the frequency observed on the screen, $\nu_m$ represents the frequency observed in a local stationary frame with the accretion disk in motion, and $\tau_k$ indicates the optical depth of the photons emitted at position $k$. Given the characteristics of the aforementioned accretion disk model, Equation (72) may reduce to (K. Wang et al. 2024)

$$I_{\nu_o} = \sum_{m=1}^{N_{\max}} f_m g^3(r_m) J_{\text{model}}(r_m). \quad (73)$$

Note that we are focusing on the radiation in the equatorial plane. The maximum number of intersections that a geodesic at the position $(\alpha, \beta)$ on the image plane can have with the equatorial plane is $N_{\max}(\alpha, \beta)$. Specifically, $N_{\max} = 1$ indicates that the geodesic intersects the equatorial plane only once, projecting a direct image of the equatorial emission onto the observer's sky. In contrast, $N_{\max} = 2, 3 \ldots$ represent geodesics that cross the equatorial plane multiple times, resulting in lensing images and higher-order images. Based on these intersections, we can calculate the radius $r_m(\alpha, \beta)$ where the geodesic intersecting the observer's imaging plane at position $(\alpha, \beta)$ crosses the equatorial plane. It is important to note that the discussion of ray trajectories is based on intersections with the equatorial plane rather than the number of turning points encountered in spacetime. For Equation (73), $J_{\mathrm{model}}(r_m)$ denotes the equatorial emissivity at $r_m$, and $g$ represents the redshift factor. Additionally, $f_m$ is the "fudge factor," which adjusts the brightness of higher-order rings (S. E. Gralla & A. Lupsasca 2020b). Variations in toy disk models can lead to changes in this parameter. For our analysis, we use $f_m = 1.5$ because the model is both geometrically and optically thin (S. E. Gralla & A. Lupsasca 2020a).

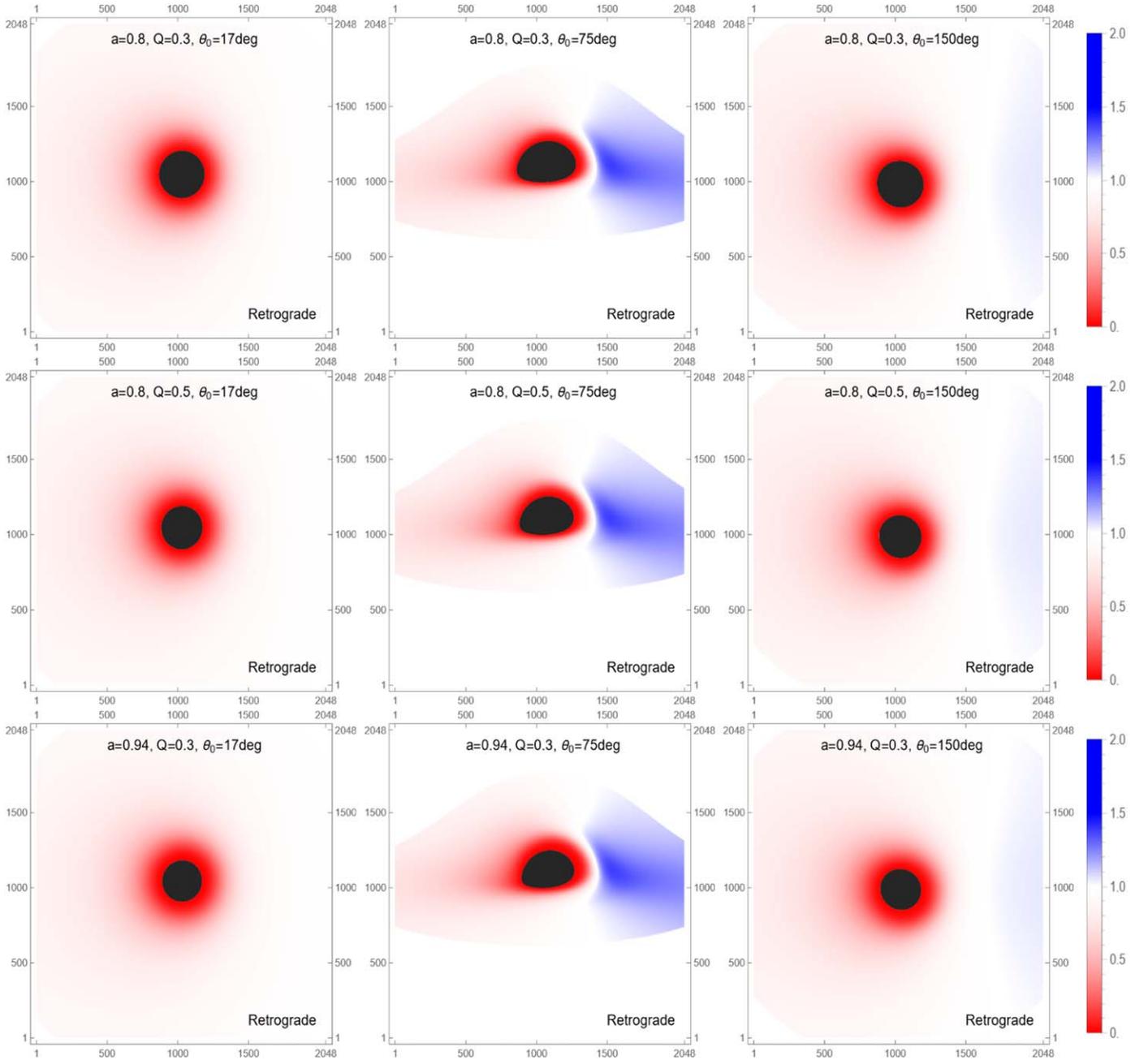

**Figure 12.** The same as Figure 10, but the disk is retrograde.

For the emissivity $J_{\mathrm{model}}(r_m)$, we follow the approach outlined by A Chael et al. (2021) and use a second-order polynomial in logarithmic space, expressed as

$$\log[J_{\mathrm{model}}(r)] = A \log[r/r_{\mathrm{H}}] + B(\log[r/r_{\mathrm{H}}])^2 . \qquad (74)$$

Considering the observation wavelengths for M87* and Sgr A* at 1.3 mm (230 GHz), we select the parameters $A = -2$ and $B = -1/2$. However, when the observation frequency decreases to 86 GHz, the values of these parameters change to $A = 0$ and $B = -3/4$, respectively (K. Akiyama et al. 2019b).

Currently, we have a method for calculating the intensity of a KN BH in a thin disk toy model. The expression for the redshift factor is given by $g = \nu_o/\nu_m$. The accretion disk model extends beyond previous models by allowing the inner edge of the accretion disk to reach the event horizon of the BH, making it reasonable to investigate a more precise redshift. Beyond the ISCO, the accretion flow continuously follows a circular orbit with an angular velocity defined by $\Omega_m(r) = (\mu^\varphi/\mu^t)|_{r=r_m}$. The impact parameter $\xi$ can be determined using Equation (7), and $\epsilon$ represents the ratio of the observed energy on the observer's screen to the conserved energy along the null geodesic, which

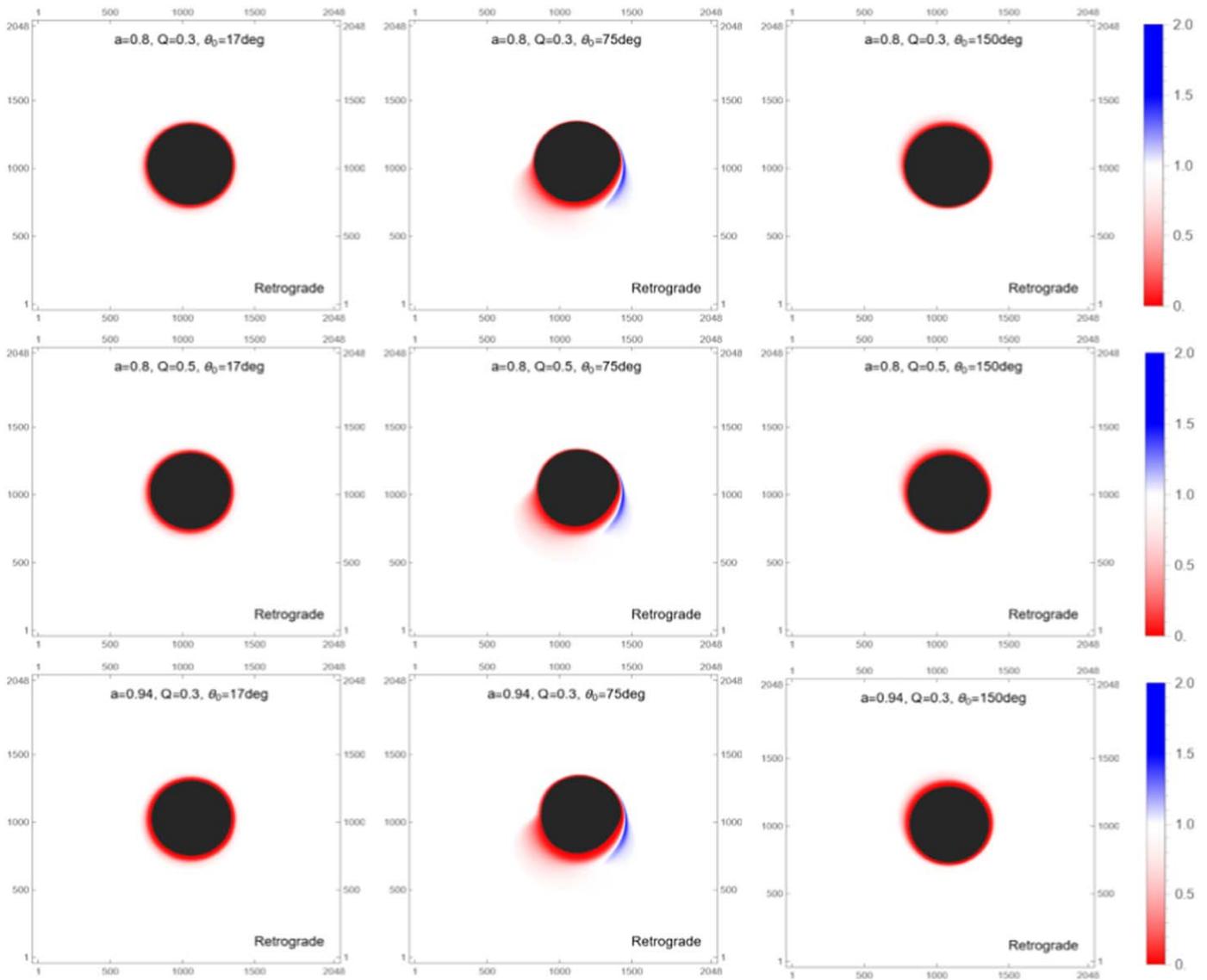

**Figure 13.** The same as Figure 11, but the disk is retrograde.

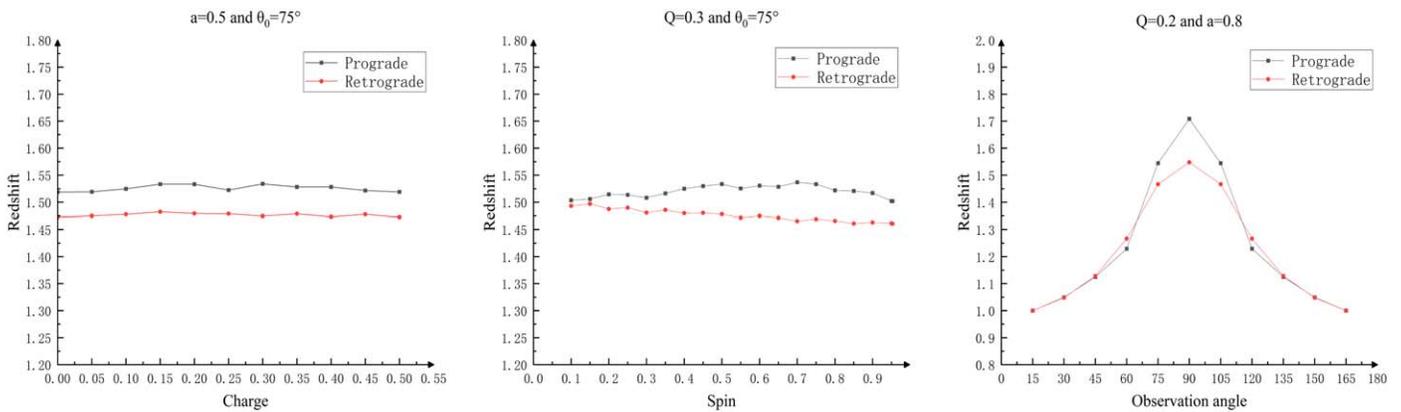

**Figure 14.** The influence of BH spin, BH charge, and observation inclination angle on the redshift (maximum blueshift point). Left panel: the BH charge varies from 0 to 0.5 with a step size of 0.05, while the spin is fixed at $a = 0.5$ and the observation angle at $\theta_0 = 75$ deg. Middle panel: the BH spin varies from 0 to 0.98 with a step size of 0.05, while the charge is fixed at $a = 0.3$ and the observation angle at $\theta_0 = 75$ deg. Right panel: the observation angle varies from 0 to 180 deg with a step size of 15 deg, while the BH charge is set to 0.2 and the spin is set to 0.8.

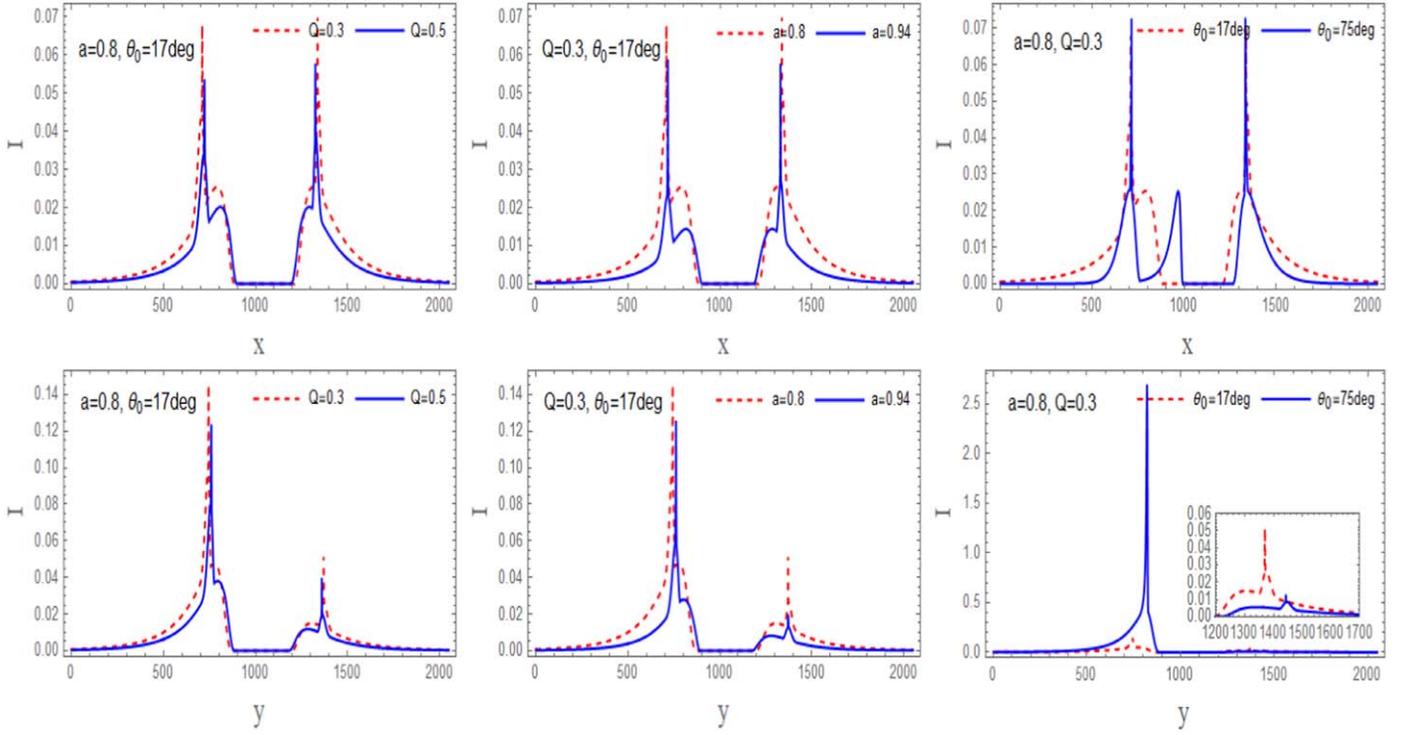

**Figure 15.** The intensity distribution of a KN BH surrounded by a prograde thin accretion disk at 230 GHz. Top panel: intensity distribution along the *x*-axis scenario. Bottom panel: intensity distribution along the *y*-axis.

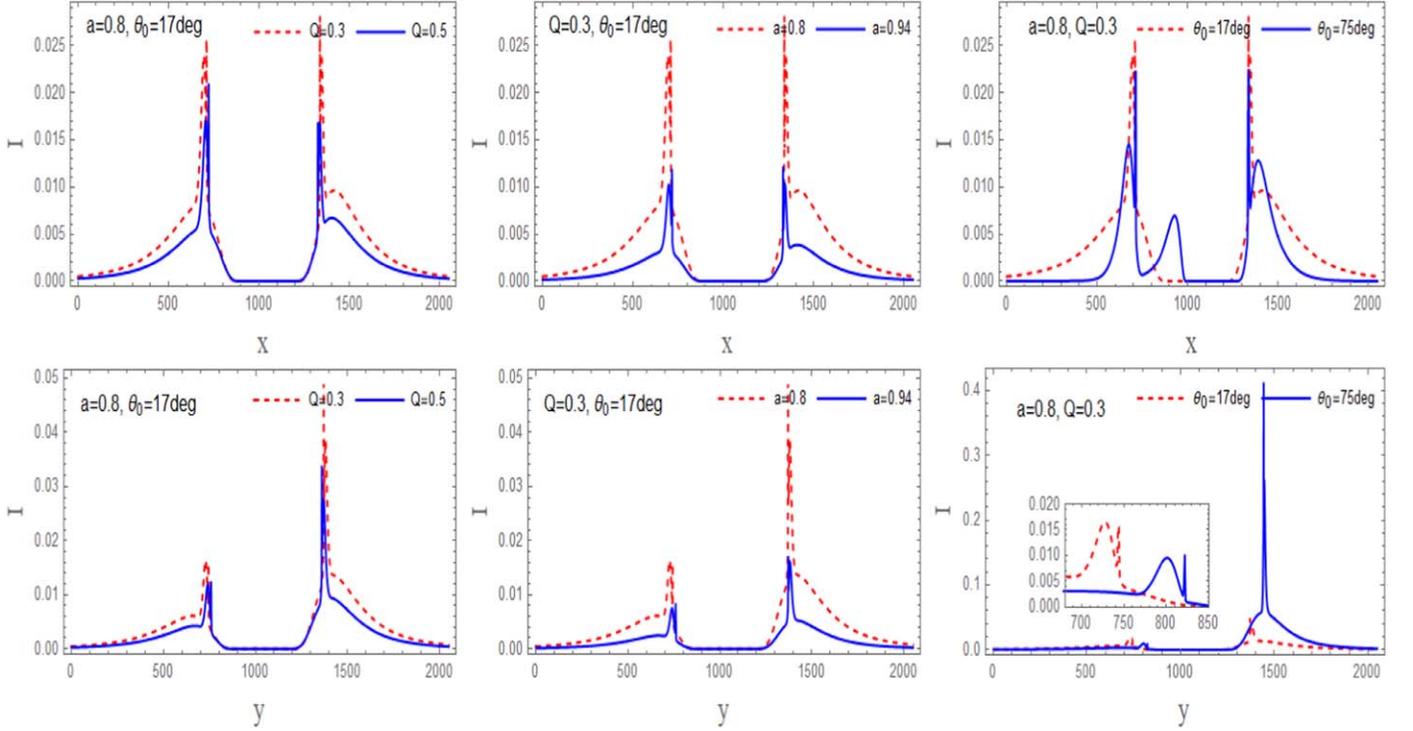

**Figure 16.** The intensity distribution of a KN BH surrounded by a retrograde thin accretion disk at 230 GHz. Top panel: intensity distribution along the *x*-axis. Bottom panel: intensity distribution along the *y*-axis.

is given by (Y. Hou et al. 2022b)

$$\epsilon = \frac{E_0}{E} = \frac{P_t}{k_t} = \delta(1 + \xi\chi), \quad (75)$$

where $\delta$ and $\chi$ are derived from Equation (64). It is important to note that $\epsilon = 1$ is valid in asymptotically flat spacetime. However, since the KN BH spacetime is not asymptotically flat, $\epsilon$ is always less than 1. Thus, when the selected position $r_m$ exceeds the ISCO, the expression for the redshift factor is

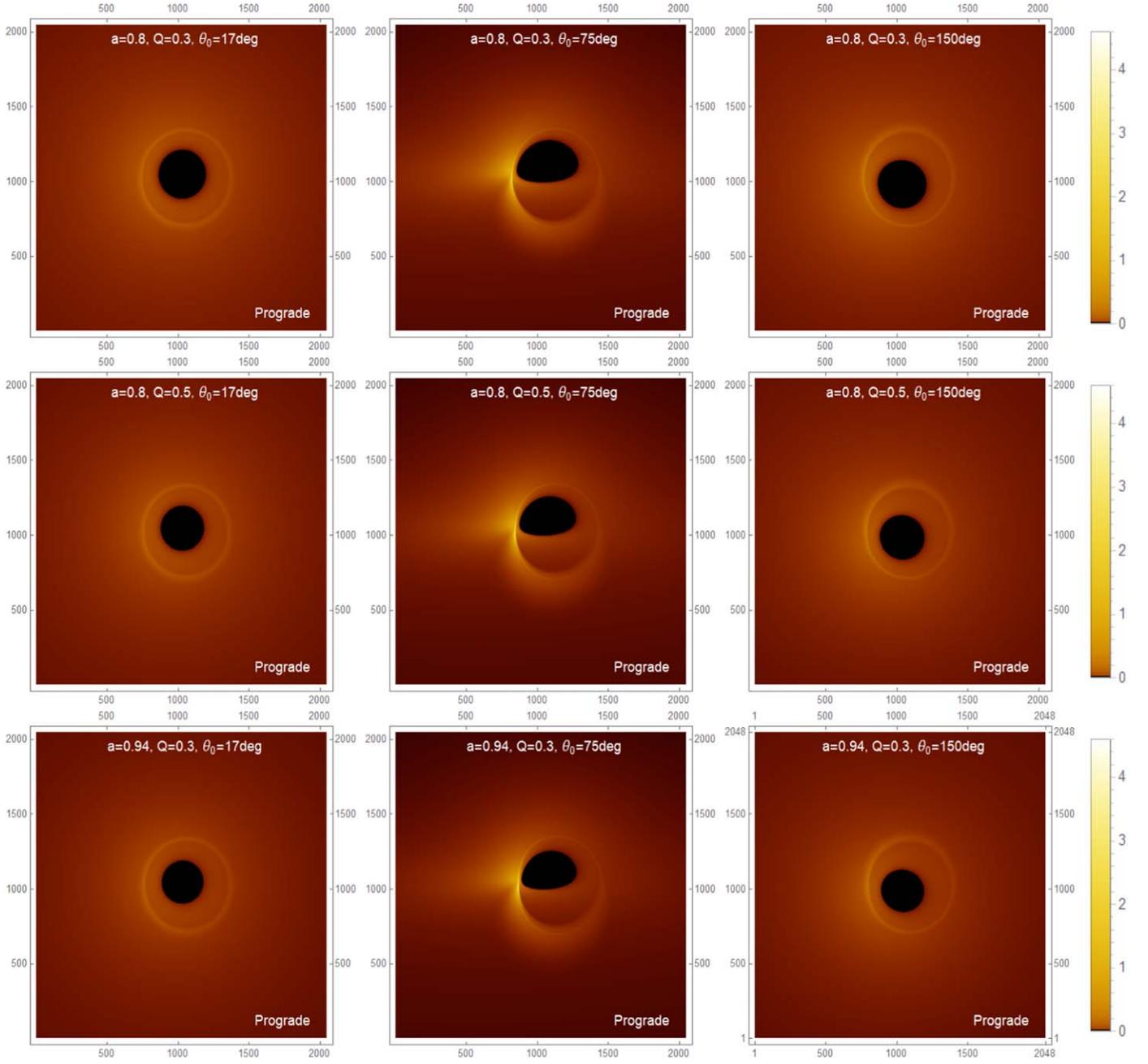

**Figure 17.** The image of the KN BH surrounded by a prograde thin accretion disk at 230 GHz. Left panel: observation angle $\theta_0 = 17$ deg. Middle panel: observation angle $\theta_0 = 75$ deg. Right panel: observation angle $\theta_0 = 150$ deg. From top to bottom, the spin of the KN BH is $a = 0.8$, $a = 0.8$, and $a = 0.94$, with a charge of $Q = 0.3$, $Q = 0.5$, and $Q = 0.3$. The mass of the BH is set to $M = 1$.

(Y. Hou et al. 2022b)

$$g_m = \frac{\epsilon}{\zeta(1 - \xi\Omega_m)}, \quad r_m > r_{\text{ISCO}}, \tag{76}$$

where we have introduced

$$\zeta = \sqrt{\frac{-1}{g_{tt} + 2g_{t\phi}\Omega_m + g_{\phi\phi}\Omega_m^2}} \Big|_{r=r_m}. \tag{77}$$

As previously discussed, the ISCO delineates the boundary of the accretion disk region. Within the ISCO region, the accretion flow follows a plunging orbit with a radial velocity $u_c^r$.

Therefore, the expression of the redshift factor is

$$g_m = \frac{\epsilon}{\frac{u_c^r k_r}{-k_t} + g^{tt}E_{\text{ISCO}} - \xi g^{t\phi}E_{\text{ISCO}} - g^{t\phi}L_{\text{ISCO}} + \xi g^{\phi\phi}L_{\text{ISCO}}}$$

$$r_m < r_{\text{ISCO}}. \tag{78}$$

To visually depict the redshift distribution, we set the fixed observation distance at $r_o = 100$, with the observer choosing $\varphi = \pi/10$ as their perspective on the screen. Considering the motion of the accretion disk, we account for two possibilities: prograde and retrograde, recognizing the presence of both forward and backward photons in the KN BH. We set the inner radius of the accretion disk at the event horizon of the KN BH, and the outer radius at $r_{\text{out}} = 20M$. Figures 10–13 present direct and lensing images that illustrate the redshift distribution of accretion disks for both prograde and retrograde scenarios. It is observed that as the observation angle increases from 0 to $\pi/2$, the region of blueshift gradually expands, particularly near the ISCO, where a more pronounced blueshift is noticeable. When the observation angle exceeds $\pi/2$ and gradually shifts toward $\pi$, the redshift range increases progressively, while the blueshift distribution steadily diminishes.

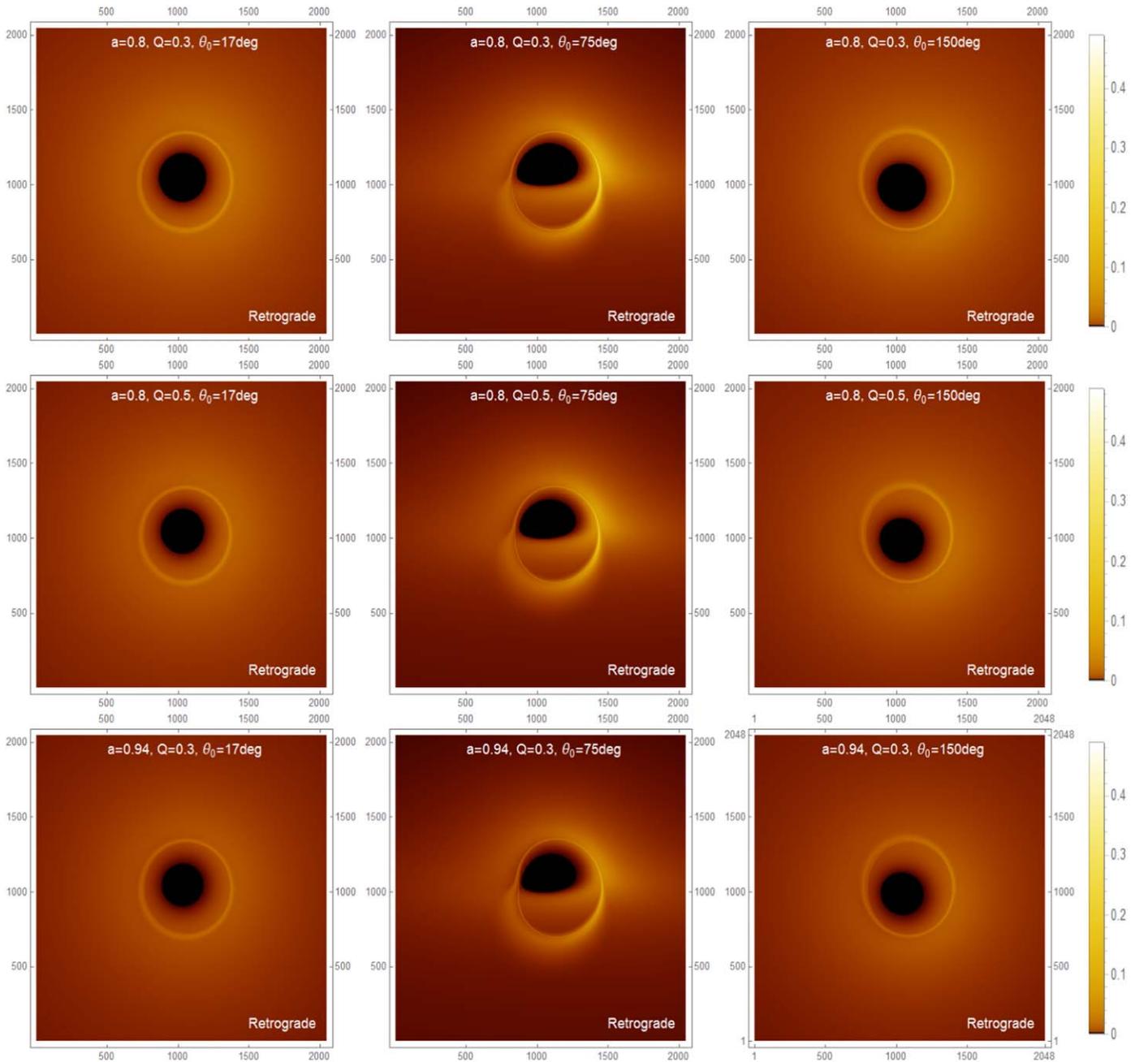

**Figure 18.** The same as Figure 17, but the disk is retrograde.

Furthermore, we plotted the effects of BH spin, charge, and observation inclination angle on redshift. As shown in the left and middle panels of Figure 14, variations in charge and BH spin have a minimal impact on the maximum blueshift point. (Note that the BH spin parameters range from $0 \leqslant a \leqslant 0.98$, and the charge range is $0 \leqslant Q \leqslant 0.5$.) In contrast, as depicted in the right panel of Figure 14, changes in the observation inclination angle have a more pronounced effect on the redshift, with the maximum blueshift point increasing as the observation inclination angle grows. From this analysis, we conclude that the influence of a BH's charge and spin on redshift distribution is relatively insignificant, with the redshift and blueshift primarily dependent on the observation inclination angle.

### 3.3. Image of the KN BH Within a Thin Disk

A detailed analysis of redshift can lead to a clear understanding of Equation (73). Next, we will concentrate on deriving the image of the KN BH. By applying Equation (73) along with fisheye camera ray-tracing technology, we can generate images of KN BHs illuminated by thin accretion disks. Given that the image of M87* was conducted at a

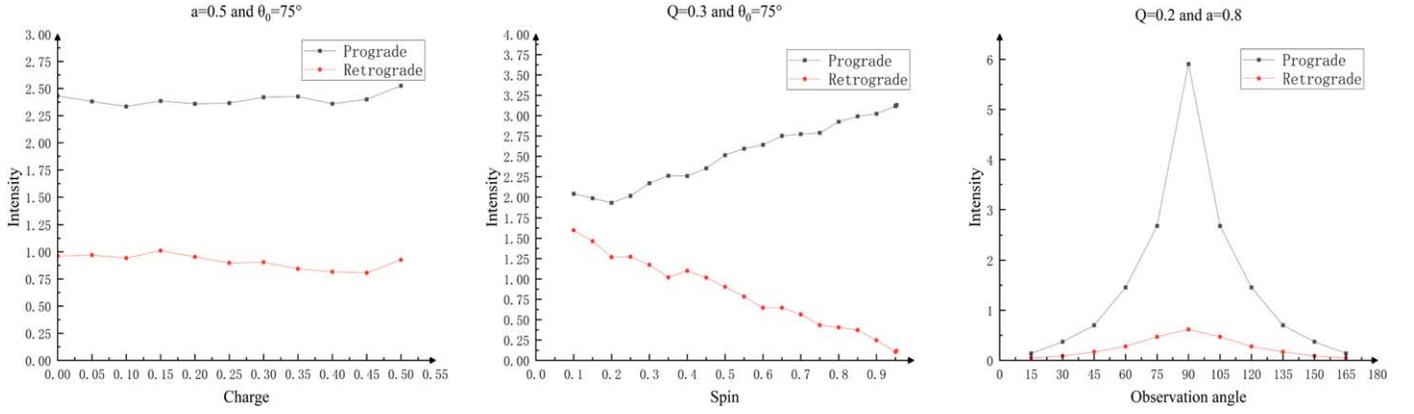

**Figure 19.** The influence of BH spin, BH charge, and observation inclination angle on the intensity. Left panel: the BH charge varies from 0 to 0.5 with a step size of 0.05, while the spin is fixed at $a = 0.5$ and the observation angle at $\theta_0 = 75$ deg. Middle panel: the BH spin varies from 0 to 0.98 with a step size of 0.05, while the charge is fixed at $a = 0.3$ and the observation angle at $\theta_0 = 75$ deg. Right panel: The observation angle varies from 0 to 180 deg with a step size of 15 deg, while the BH charge is set to 0.2 and the spin is set to 0.8.

frequency of 230 GHz, we choose to use Equation (74) as the appropriate formula:

$$\log[J_{\text{model}}(r)] = -2\log[r/r_H] - \frac{1}{2}(\log[r/r_H])^2. \quad (79)$$

To gain a more comprehensive understanding of the optical appearance of the KN BH, we have also derived the results at 86 GHz, enabling comparison with the observations made at 230 GHz. Based on Equations (73) and (79), we initially present the intensity distribution in the $x$- and $y$-directions. Figures 15 and 16 illustrate the intensity distribution along the $x$-axis and $y$-axis on the screen for both prograde and retrograde accretion disks. In the $x$-direction (first column), regardless of whether the accretion disk is prograde or retrograde, an increase in charge leads to a decrease in peak intensity, with the concave area of the intensity distribution contracting inward. Additionally, an increase in angular momentum results in a decrease in peak intensity, particularly noticeable for retrograde accretion disks. The BH's spin further causes the central intensity depression area to contract inward. The effects of changes in the observation inclination angle vary: for a prograde accretion disk, an increase in inclination angle results in an increase in peak intensity, and the central intensity depression area expands outward. In contrast, for retrograde accretion disks, peak intensity decreases as the inclination angle increases.

In the intensity distribution along the $y$-direction (second column), the prograde and retrograde accretion disks display completely opposite behaviors in terms of the positions where the maximum intensity occurs. However, the effects of BH charge, spin, and observed inclination angle remain consistent with those observed in the $x$-direction. As the inclination angle increases, the intensity distribution along the $y$-direction shows a rapid rise. Figures 17 and 18 present images of a KN BH surrounded by a thin accretion disk, illustrating both prograde and retrograde accretion disks. It is clear that, regardless of the accretion disk's orientation, an increase in the observed inclination angle $(0–\pi/2)$ allows for a clear distinction between direct and lensing images. However, as the observation angle approaches $\theta_0 = 150$ deg, the brightness distribution near the image becomes less distinct, making it increasingly difficult to differentiate between direct images and lensing images.

Furthermore, the spin of BHs significantly influences the shape of inner shadows, while the effect of charge, though present, is comparatively weak. Regardless of changes in the BH's spin and charge, at both low and high observation angles, internal shadows and critical curves remain discernible, indicating that they are intrinsic spacetime features of the BH. Additionally, the central brightness depression is a characteristic phenomenon observed in all cases where the accretion disk is not excessively thick. The Doppler effect caused by prograde and retrograde motion is clearly visible on opposite sides of the screen. Notably, while the Doppler is more pronounced on the right side of the imaging screen in the case of a retrograde accretion disk, the brightness of the KN BH decreases due to the drag effect of its spin.

Figure 19 illustrates the effects of BH charge, spin, and observation inclination on intensity (including both prograde and retrograde accretion disks). The left panel shows that the influence of charge on intensity is weak (note that the charge range is $0 \leqslant Q \leqslant 0.5$). The middle panel demonstrates the effect of BH spin on intensity: for a prograde accretion disk, intensity increases steadily with increasing BH spin, while for a retrograde accretion disk, the effect is the opposite-intensity gradually decreases as spin increases. Regarding the influence of the observation angle, within the range of $(0–\pi/2)$, the maximum intensity rises rapidly with the increase of the observation angle, reaching its peak at $\theta_0 = \pi/2$, and then gradually decreases as the observation angle continues to increase.

In addition to deriving images at 230 GHz, we also examined images of KN BHs surrounded by thin accretion disks at 86 GHz for comparative analysis. In this scenario, Equation (74) simplifies to

$$\log[J_{\text{model}}(r)] = -\frac{3}{4}(\log[r/r_H])^2. \quad (80)$$

Figure 20 shows the intensity distribution in the $x$- and $y$-directions of a KN BH surrounded by a thin accretion disk at 86 GHz. Compared to the 230 GHz scenario, the overall intensity has increased, and the position of the maximum peak has shifted, regardless of whether the accretion disk is prograde or retrograde. The effects of charge, BH spin, and observation inclination angle on the intensity distribution remain largely consistent with those observed at 230 GHz.

Finally, we generated an image of a KN BH illuminated by a thin accretion disk at 86 GHz for comparison with the 230 GHz scenario (Figure 21). It is evident that the brightness of the radiation in the field of view is influenced by gravitational redshift, while the central brightness depression, or inner shadow, remains present. Additionally, the critical curve remains unchanged, indicating that the inner shadow is an inherent spatiotemporal characteristic of BHs.

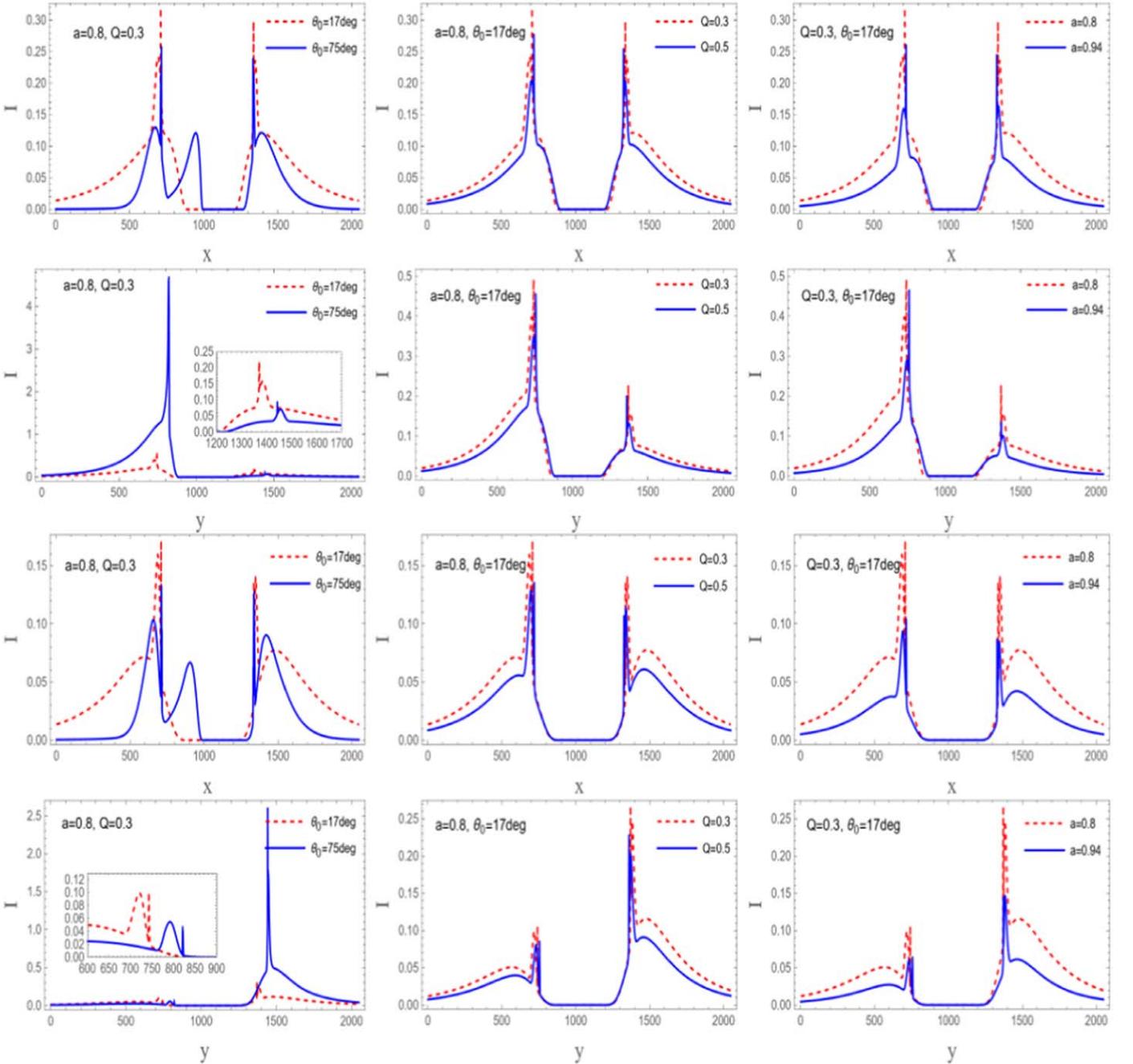

**Figure 20.** The intensity distribution of a KN BH surrounded by a prograde and retrograde thin accretion disk at 86 GHz. Top panel: intensity distribution along the *x*-axis for the prograde disk. Middle top panel: intensity distribution along the *y*-axis scenario for the prograde disk. Middle bottom panel: intensity distribution along the *x*-axis for the retrograde disk. Bottom panel: intensity distribution along the *y*-axis for the retrograde disk.

## 4. Conclusions and Discussion

In this analysis, we investigated the image of the KN BH surrounded by a thin disk. Initially, we explored the dynamics of the KN BH and examined photon trajectories around it using ray-tracing techniques. By introducing elliptic integrals, we analyzed the conditions under which photons can orbit the KN BH multiple times. To determine the shadow of KN BH, we derived a fourth-order function related to the radial coordinate $\tilde{r}$ and used numerical methods to find the position of the critical curve. The results indicate that under varying BH charges, the radial equation yields four distinct roots, each satisfying specific inequalities associated with the event horizon.

For the case of a KN BH surrounded by an equatorial accretion disk, we employed elliptic integrals to derive contour lines at various radial positions. It is observed that increasing the observation inclination angle causes the critical curve to expand outward. Additionally, at smaller observation inclination angles, the secondary image of particle orbits remains within the primary image, similar to the behavior observed for spherically symmetric BHs. However, as the observation inclination angle increases, the primary and secondary images separate and form a cap-like structure. The results also indicate that the BH's spin and the observation inclination angle significantly affect the asymmetry of the image and the distortion of the inner shadow, while variations in charge have minimal impact on these deformations. To visually demonstrate how light is influenced by the strong gravitational field of the KN BH, we divided the observation plane into 50 regions and employed ray-tracing techniques to illustrate the light trajectories around the KN BH (Figure 4).

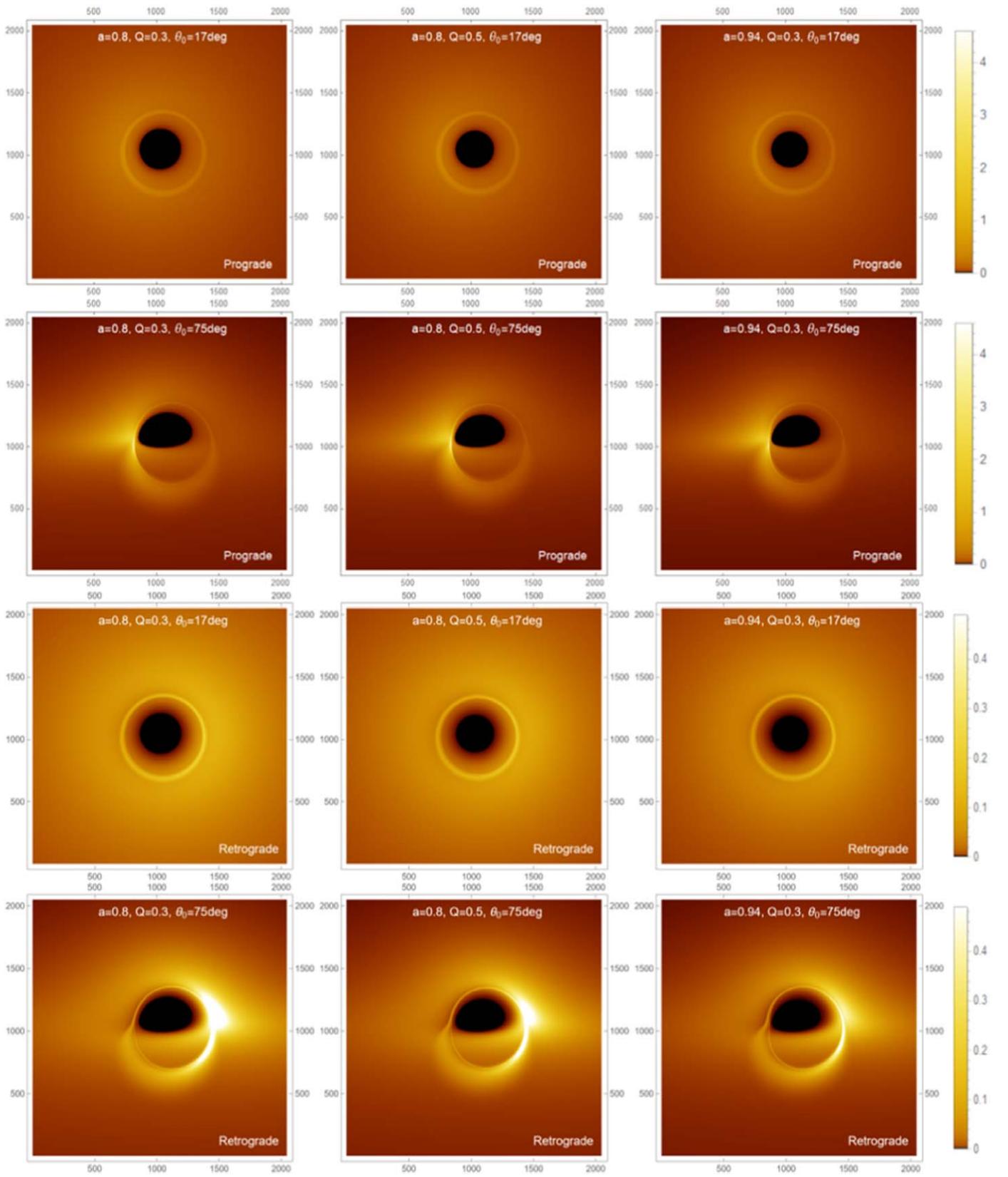

**Figure 21.** The image of the KN BH surrounded by a prograde and retrograde thin accretion disk at 86 GHz. Top panel: observation angle $\theta_0 = 17$ deg for the prograde disk. Middle top panel: observation angle $\theta_0 = 75$ deg for the prograde disk. Middle bottom panel: observation angle $\theta_0 = 17$ deg for the retrograde disk. Bottom panel: observation angle $\theta_0 = 75$ deg for the retrograde disk. From left to right, the spin of the KN BH is $a = 0.8$, $a = 0.8$, and $a = 0.94$, with corresponding charge of $Q = 0.3$, $Q = 0.5$, and $Q = 0.3$. The mass of the BH is set to $M = 1$.

To investigate the optical appearance of the KN BH, we extended the classical accretion disk model to a scenario where the innermost region of the accretion disk extends to the BH's event horizon. In this context, it was necessary to reevaluate the redshift distribution and radiation transfer within the ISCO, as well as to conduct more extensive ray-tracing calculations over a broader integration region. Consequently, we rederived the effective potential function and redshift expression, finding that the redshift distribution of the accretion disk depends on the BH's charge, spin, and observation inclination angle. Meanwhile, the prograde and retrograde disk exhibited distinct redshift characteristics: for a prograde disk, blueshift appears on the left side of the accretion disk, while for a retrograde disk, blueshift appears on the right side. We also examined the effects of BH charge, spin, and observation inclination on redshift. The results indicated that charge and spin had minimal impact on the redshift (maximum blueshift point). However, as the observation inclination angle increased from 0 to 90 deg, the redshift gradually increased. Beyond 90 deg, the redshift then gradually decreased.

By considering a zero-angular-momentum observer, we developed a model for light-intensity conservation along geodesics. Using fisheye camera ray-tracing techniques, we conducted a detailed analysis of the image of a thin accretion disk surrounding a KN BH. The results provided insights into the optical appearance and intensity distribution of the KN BH at 230 GHz and 86 GHz observation frequencies. We observed notable differences between the images of prograde and retrograde accretion disks. Particularly, higher observation inclination angles accentuated the boundary region between direct images and lensed images. The results indicate that an increase in charge results in greater light bending, while spin leads to deformation of the inner shadow. Comparing observations at two frequencies, we found that both the total and peak intensity at 86 GHz are higher than at 230 GHz, while spatiotemporal features, such as the inner shadow and critical curve, remain unchanged. Interestingly, both real astronomical observations and theoretical predictions consistently show a bright outer ring surrounding a central brightness depression.

## Acknowledgments

We are deeply grateful for the insightful comments provided by the referee, as well as the effective discussions with Professor Minyong Guo from Beijing Normal University, Dr. Zhenyu Zhang, Dr. Yehui Hou, and Dr. Jiewei Huang from Peking University. This work is supported by the National Natural Science Foundation of China (grant Nos. 12133003, 42230207) and the Fundamental Research Funds for the Central Universities, China University of Geosciences (Wuhan; grant No. G1323523064).

## ORCID iDs

Sen Guo https://orcid.org/0000-0001-6802-9860
En-Wei Liang https://orcid.org/0000-0002-7044-733X